\documentclass[
showkeys,12pt, prd,
preprint,preprintnumbers,nofootinbib,
groupedaddress,superscriptaddress,amsmath,amssymb]{revtex4}
\usepackage{graphicx}
\usepackage{dcolumn}
\usepackage{bm}
\usepackage{amssymb}
\usepackage{amsmath}
\usepackage{epsfig}    
\usepackage{color}
\usepackage{slashed}
\usepackage{hhline}

\def\be{\begin{equation}}
\def\ee{\end{equation}}
\newcommand{\bea}{\begin{eqnarray}}
\newcommand{\eea}{\end{eqnarray}}
\newcommand{\nn}{\nonumber}

\numberwithin{equation}{section}

\begin{document}


\title{Radiative neutrino  model with semi-annihilation  dark matter}

\author{Haiying Cai}
\email{haiying.cai@apctp.org}
\affiliation{Asia Pacific Center for Theoretical Physics, Pohang, Gyeongbuk 790-784, Republic of Korea}

\date{\today}

\begin{abstract}
We propose a Two-Loop induced radiative neutrino model with hidden gauged $U(1)$ symmetry, in which  a dark matter  of Dirac fermion arises.  The  relic density gets   contribution  from annihilation and  semi-annihilation  due to a residual  $\mathbb{Z}_3$  parity.  After imposing the requirement of neutrino oscillation data and lepton flavour violation bounds,  we find out that the semi-annihilation plays a crucial role in order to satisfy the relic density  constraint $0.117 < \Omega h^2 < 0.123$,  by proceeding near either one of two deconstructive scalar resonances. Our numerical analysis  demonstrates the  allowed region for the DM-Scalar coupling with  the DM mass in $(80, 400)$ GeV.  
\end{abstract}
\maketitle
\newpage

\section{Introduction}
Radiative  seesaw neutrino  models are one of the attractive scenarios to connect neutrino sector with dark matter (DM) sector in a natural manner. These two sectors certainly involve mysterious puzzles that are frequently interpreted as  physics beyond the Standard Model (SM). When the neutrino masses are radiatively induced, the magnitude of relevant  couplings  could reach $\mathcal{O}(1)$  compared with the case where the neutrino mass is generated at the tree-level, so that  the mass hierarchy  among the SM sector and  heavy fermion/scalar sectors  is largely alleviated. Furthermore, new particles that are accommodated in the theory  are  at $\mathcal{O}$(TeV) energy scale  and   accessible by the extensive search at Large Hadron Collider (LHC). For radiative seesaw,  a discrete symmetry is essentially implemented  in order to forbid  the neutrino mass  at the tree-level and such symmetry  will  in turn  stabilize the lightest neutral particle  as a DM candidate.  As a consequence, this type of theory provides  interesting  phenomenologies  with the requirement to satisfy the observed relic density of  $\Omega h^2\approx0.120 \pm 0.001$~\cite{planck} and other experimental constraints.

The simplest discrete symmetry can be $\mathbb{Z}_{2}$,  as the remnant of  a broken $U(1)$  symmetry, and a typical DM-generated neutrino mass model at the one-loop level is proposed in~\cite{Ma:2006km}. However  other  enlarged discrete symmetry is also possible to stablize the DM candidate such as the $\mathbb{Z}_N$, $N>2$ discrete parity, which brings in semi-annihilation in addition to  annihilation for the  Lee-Weinberg scenario~\cite{Lee:1977ua}, allowing for  an odd number of DM particles appearing in a $2\to2$ process~\cite{Hambye:2008bq,Hambye:2009fg,DEramo:2010keq, Belanger:2012vp}. Under the control of  $\mathbb{Z}_N$  discrete symmetry,  any  field transforming as  $f_i \to \omega^a f_i$, with $\omega = \exp(i 2 \pi /N)$ and $a = 1, \cdots, N-1$, could serve as the dark matter candidate depending on the spectrum and interactions.  In this paper we consider a two-loop induced neutrino mass model~\cite{2-lp-zB, Babu:2002uu, Ma:2007gq, Kajiyama:2013zla, Kajiyama:2013rla, Aoki:2013gzs} with new particles charged under a hidden $U(1)$ symmetry~\cite{Langacker:2008yv, Ma:2013yga,  Ko:2014loa, Ma:2015mjd, Chun:2008by, Ko:2016ala, Nomura:2018jkd}, in which a Dirac fermion type of $\mathbb{Z}_3$ DM candidate arises,   whose relic density is dominantly explained by the s-channel of semi-annihilation modes.  Note that in this model, a complex scalar is also possible to behave as DM in the inverse mass pattern. The discrete $\mathbb{Z}_3$ symmetry origins from  the spontaneous breaking of $U(1)$ symmetry and plays an important role to ensure the DM $\chi$ does not decay  while the reaction in the form of $\chi \chi \to \chi^\dagger v_i$ exits. We present how the DM and  neutrino mass are correlated  by formulating  each sector. In particular, we  perform an analysis to obtain the allowed region which satisfies a set of  necessary bounds,   including neutrino oscillation data, Lepton Flavour Violations (LFVs), muon anomalous magnetic moment ($\Delta a_\mu$, {\it{aka}} muon $g-2$), and the DM relic density.

This paper is organized as follows.
In Sec.~II, we show the valid Lagrangian with charge assignments, and formulate the scalar and neutrino sectors, along with the LFVs,  muon $g-2$, $Z-Z'$ mixing and bound of electroweak precision test. In Sec.~III, we analyze the Dirac fermionic DM  to explain the relic density  with an emphasis on the semi-annihilation  and a brief illustration of the analytic derivation.   In Sec.~IV, we conduct a numerical analysis, and show the allowed region to satisfy all the phenomenologies that we discuss above.
Finally we conclude and discuss in Sec.~V.

\section{ The Model} \label{sec:model}
\vspace{-15pt}

 \begin{widetext}
\begin{center} 
\begin{table}
\begin{tabular}{|c||c|c|c|c|c||c|c|c|c||c|c|c|c|}\hline\hline  
&\multicolumn{5}{c||}{Fermion Fields} & \multicolumn{4}{c||}{Scalar Fields}& \multicolumn{4}{c|}{Inert Scalar Fields} \\\hline
& ~$L_L$~ & ~$e_R^{}$~ & ~$L'_{L/R}$~ & ~$\chi_{L/R}$~ & ~$N_{L/R}$ ~ & ~$H$~&~ $H'$~  & ~$\Delta$~ & ~$\varphi$
~ & ~$s$~ & ~$\eta$~ & ~$s'$~ & ~$\eta'$ \\\hline 
$SU(2)_L$ & $\bm{2}$  & $\bm{1}$  & $\bm{2}$ & $\bm{1}$ & $\bm{1}$ & $\bm{2}$& $\bm{2}$ & $\bm{3}$& $\bm{1}$& $\bm{1}$& $\bm{2}$& $\bm{1}$& $\bm{2}$ \\\hline 
$U(1)_Y$ & $-\frac12$ & $-1$  & $-\frac{1}{2}$ & $0$ & $0$ & $\frac{1}{2}$& $\frac{1}{2}$& $1$&  {0} & {0} & $\frac{1}{2}$& {0} & $\frac{1}{2}$ \\\hline
$U(1)_H$ & $0$ & $0$  & $2x$ & $x$ & $y$ & $0$ & $-3x$ & $-3x$& $-3x$ & $-2x$ & $x$ & ${x+y}$ & $-2x+y$ \\\hline
\end{tabular}
\caption{Contents of fermion and scalar fields
and their charge assignments under $SU(2)_L\times U(1)_Y\times U(1)_H$, where all the new fields are singlets under $SU(3)_C$, and all the quark fields are neutral under $U(1)_H$. Note that the $H'$  field can only present for the   one-loop radiative seesaw. }
\label{tab:1}
\end{table}
\end{center}
\end{widetext}

\begin{figure}[tb]
\begin{center}
\includegraphics[scale=0.56]{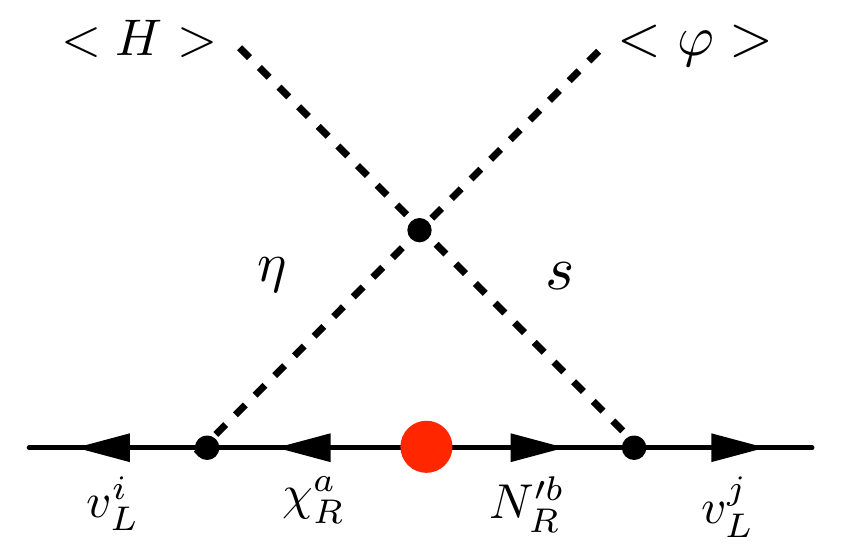}
\qquad \quad \quad
\includegraphics[scale=0.56]{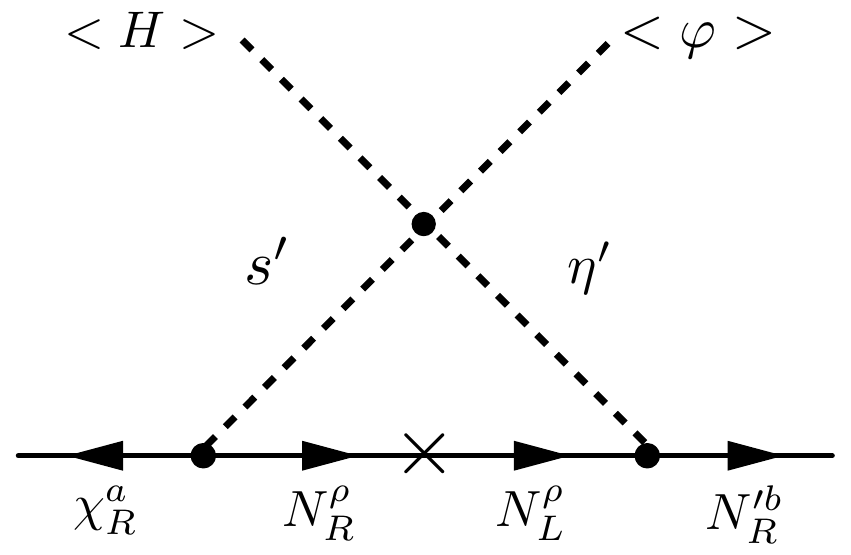}
\caption{Neutrino mass in the gauge basis at the two-loop level, where the right plot represents the red dot in the left plot.}   \label{fig:2loop}
\end{center}\end{figure}

The model is built by extending the SM with additional scalars and vector-like fermions, which are charged under a hidden $U(1)$ symmetry before some of the scalars obtain VEVs. The field contents and their charge assignments are reported in Table~\ref{tab:1}. For the fermion sector, an isospin doublet   $L'\equiv[E',N']^T_i$  plus two isospin singlets $\chi_i$ and $N_i$ with $i = 1,2, 3$, are added.  The vector-like  nature of these extra fermions ensures our extension to be anomaly-free.  The quantum number assignment for $L^\prime$, $\chi$, $N$ under the two gauge groups of $(U(1)_Y, U(1)_H)$ are  $(-1/2,2x)$, $(0, x)$ and $(0,y)$  respectively. Here we use two arbitrary integers $(x,y)$  with   $\{x,y\} \neq 0$ to keep track of the heavy fermions running in the outer and  inner loops of neutrino mass diagram (see Figure~\ref{fig:2loop}).
As for new scalar fields, we introduce four inert scalar fields $s$, $\eta$, $s'$, $\eta'$, where $(\eta, \eta')$ are  $SU(2)_L$ doublets and $(s, s')$ are singlets. As we can see that  since $(s, \eta)$ are charged under the $U(1)_H$ as $(-2x, x)$,  these two fields will only interact with new fermions of $L'$ and $\chi$.  On the other hand,  the two prime fields  $(s', \eta')$  are  charged with $(x+y, -2x+y)$ for the hidden symmetry, thus they are allowed to connect with the exotic fermion $N$ under the assumption of $y \neq- x$, $y\neq 2x$~\footnote{In fact we can think that the gauged $U(1)_H$ is a linear combination of two global $U(1)$s, which should be observed individually in the unbroken phase.}. The  two scalar fields $(H, \varphi)$ are needed in order to generate the neutrino mass at the two-loop level provided they will mix inert scalars $(\eta, s)$ and $(\eta' , s')$ inside each set. The symmetry permits more scalar fields, like a doublet $H'$ or a triplet $\Delta$, to induce the $Z$-$Z'$ mixing for  LHC collider signature.  In case of adding $H'$, the neutrino mass will be generated at the one-loop level,  since the red dot in the  Figure.\ref{fig:2loop}  can be substituted by an interaction of $ \overline{L^{\prime c}}_{L/R} H' \chi_{L/R}$. However in such case the VEV of $H'$ should be very small (equivalent to  loop generated), so that this one-loop radiative seesaw is possible to reconcile the tension between neutrino mass and relic density bound.  Thus we will  focus on  exploring the impact of a triplet $\Delta$ interplaying with $(H, \varphi)$ via the scalar potential in the two-loop radiative seesaw. For that purpose, the scalars $H$, $\Delta$  and $\varphi$ are required to develop nonzero vacuum expectation values (VEVs), respectively symbolized by $\langle H\rangle\equiv v_H/\sqrt{2}$, $\langle \Delta \rangle\equiv v_{\Delta}/\sqrt{2}$, $\langle \varphi\rangle\equiv v_\varphi/\sqrt{2}$. The valid renormalizable Lagrangian  for the fermion sector are given by,
\begin{align}
-\mathcal{L}_{Y}
&=
y_{\ell_{ii}} \bar L_{L_i} H e_{R_i}  + y_{\chi_{ab}}^{L} s \bar\chi^c_{L_a}\chi_{L_b}+ y_{\chi_{ab}}^{R} s \bar\chi^c_{R_a}\chi_{R_b} + y_{\eta_{ia}} \bar L_{L_i}\tilde\eta\chi_{R_a} + y_{S_{ia}} s \bar L_{L_i}L'_{R_a} \nn\\
&
 + y_{\eta'_{ab}} \bar L'_{R_a}\tilde\eta' N_{L_b} + y'_{\eta'_{ab}} \bar L'_{L_a}\tilde\eta' N_{R_b} 
+ y_{s'_{ab}} \bar N_{R_a}\chi^c_{R_b} s'  + y'_{s'_{ab}} \bar N_{L_a}\chi^c_{L_b} s'  \nn\\
& + M_{\chi_{aa}}\bar\chi_{L_a} \chi_{R_a} + {M_{N_{aa}}\bar N_{L_a} N_{R_a}}
+ M_{N'_{aa}}\bar L'_{L_a} L'_{R_a}
+ {\rm h.c.}, 
\label{Eq:lag-flavor}
\end{align}
where $i, a,b=1,2, 3$ are the  flavor indices for the SM and exotic fermions, and $\tilde \eta\equiv i\sigma_2\eta^*$, with $\sigma_2$ being the second Pauli matrix.  For simplicity, we assume that all coefficients are real and  $M_\chi, M_{N}, M_{N'}$ to be diagonal matrices. The first term of $\mathcal{L}_{Y}$ generates the SM charged-lepton masses $m_{\ell_i} \equiv y_{\ell_i} v_H/\sqrt2$,  while the $2$nd to $4$th terms will be responsible  for the (semi-)annihilations.  The residual $\mathbb{Z}_3$  from the broken hidden  symmetry makes the lightest neutral states with  $U(1)_H $ charge of $x$ or  $\pm 2x$, i.e. particles in the set of $(\chi_i , N'_i, \eta, s)$,  to be our DM candidate.  While in this paper we are interested in the mass pattern where $\chi_1$ actually plays the role of   DM.  Referring to Table~\ref{tab:1}, we can see that the two scalar fields $( \Delta, \varphi)$ carrying a $U(1)_H$ charge $q_H = -3x$ with $x \in $ integer, so that they will transform under the Abelian $U(1)$ symmetry as $\Delta \to e^{- i q_H \alpha} \Delta $ and $\varphi \to e^{- i q_H \alpha} \varphi $, for an arbitrary value of $\alpha$ before the spontaneous symmetry breaking.  However after these two scalars  obtain VEVs,  the phase  is forced  to be  $\alpha = 2 \pi/ 3 $ for any integer $x\geq1$,  thus  the Lagrangian is still invariant under a discrete $\mathbb{Z}_3$ symmetry. And the particles with $x (2x)$ charge in $U(1)_H$  have $w=e^{i 2\pi/3} (w^2)$ parity assignment under this $\mathbb{Z}_3$.

\subsection{ The scalar potential}
We explicitly write the nontrivial terms for the inert scalar potential  which are invariant under the  $SU(2)_L\times U(1)_Y\times U(1)_H$ gauge symmetry to be:
\begin{align}
\mathcal{V}_1&= 
\left(\lambda_0 H^\dag\eta s^*\varphi 
+{ \lambda'_0 H^\dag\eta' s^{\prime*}\varphi^*  } +{\rm h.c.}\right)  + \sum_{\phi}^{\eta,\eta', s, s'} \left[\lambda_{H\phi} (H^\dag H) (\phi^\dag \phi )+ \lambda_{\varphi \phi} (\varphi^\dag \varphi) (\phi^\dag \phi) \right]  \nonumber \\ &  + \sum_{\phi}^{\eta,\eta'} \lambda'_{H\phi} (H^\dag \phi) (\phi^\dag H ) 
+  \sum_{\phi}^{\eta,\eta', s, s'} \left[\mu^2_\phi \phi^\dag \phi +  \lambda_\phi |\phi^\dag \phi|^2 \right] \,.
\label{Eq:pot1}
\end{align}
where we assume that these terms like $s^{\prime 2}\varphi^{(*)}$, $s^{\prime 2}\varphi^{2}$,  $\eta^{\prime \dag}\eta \varphi^2$ vanish  due to $U(1)_H$ charges (e.g. $x= y =1$). Thus no mass splitting occurs among the real and imaginary parts of any inert field.    The general  potential  for the scalars $(H, \Delta)$ can be found in ref~\cite{Bonilla:2015eha, Primulando:2019evb},  and we will modify it by adding  interactions with a complex singlet $\varphi$. 
\begin{eqnarray}
 \mathcal{V}_2 &=& -\mu_H^2 H^\dagger H + \lambda_H ( H^\dagger H)^2 + M^2 Tr(\Delta^\dagger \Delta) +\lambda_1 (Tr (\Delta^\dagger \Delta) )^2  \nn \\ 
& +&  \lambda_2 Tr((\Delta^\dagger \Delta)^2 )  + \lambda_3 (H^\dagger H) Tr (\Delta^\dagger \Delta ) + \lambda_4 H^\dagger \Delta \Delta^\dagger H \nn \\ & - & \mu_\varphi^2 \varphi^*\varphi + \lambda_\varphi (\varphi^* \varphi)^2 + \lambda_5 H^\dag H  \varphi^* \varphi   +  \lambda_6 Tr (\Delta^\dag \Delta)   \varphi^* \varphi  \nn \\  &+& \left[ \lambda_\Delta H^T i \sigma_2 \Delta^+ H \varphi
+{\rm h.c.}\right]
\label{Eq:pot2}
\end{eqnarray}
The scalar  fields beside the inert ones  are explicitly expressed as: 
\begin{align}
H = \left[\begin{array}{c} G^+ \\  \frac{v_H+ h+ i G^0}{\sqrt{2}} \end{array} \right] \,, \quad 
\Delta=\left[
\begin{array}{cc}
\frac{\Delta^{+} }{\sqrt{2}} & \Delta^{++} \\
\frac{v_\Delta + \Delta_R + i \Delta_I}{\sqrt{2}} & - \frac{\Delta^+}{\sqrt{2}}
\end{array}\right]\,, \quad 
\varphi = \frac{v_\varphi+\varphi_R+i \varphi_I}{\sqrt2} \,.
\end{align}
so that the mass of $W$ boson is fixed to be  $m_{W} = \frac{g_2 \sqrt{v_H^2+ 2v_{\Delta}^2}}{2}   $.
The minimum of the potential is determined  by derivatives  $\partial\mathcal{V}_2/\partial v_H=0$, $\partial\mathcal{V}_2/\partial v_\Delta=0$,
$\partial\mathcal{V}_2/\partial v_\varphi=0$, which read as:
\begin{eqnarray}
 & &-\mu_H^2 +  \lambda_H v_H^2 + \frac{\lambda_3+\lambda_4}{2} v_\Delta^2 + \frac{\lambda_5 } {2} v_{\varphi}^2 =  \lambda_\Delta v_\Delta v_\varphi   \nn \\
 & &  (M^2 +  \frac{\lambda_3 +\lambda_4}{2} v_H^2 + (\lambda_1+\lambda_2) v_\Delta^2 + \frac{\lambda_6}{2} v_{\varphi}^2) v_\Delta =  \lambda_\Delta v_H^2 v_\varphi /2  \nn \\
& & ( - \mu_\varphi^2 + \lambda_\varphi v_\varphi^2  +  \frac{\lambda_5 } {2} v_{H}^2 + + \frac{\lambda_6 } {2} v_{\Delta}^2) v_\varphi = \lambda_\Delta v_H^2 v_\Delta /2 
\end{eqnarray}
As we argue in the section~[\ref{sec:zzH}] for $Z-Z'$ mixing, $v_\Delta$ is very tiny due to the  $\rho$ parameter, thus we will focus on the limit of $v_\Delta \ll v_H \lesssim v_\varphi$.  Thus under the assumption of negligible mixing between $\varphi$ and  $(H, \Delta)$, i.e. $\lambda_{5}, \lambda_6 \ll 1$, we obtain:
\begin{align}
 v_H \simeq \left(\frac{\lambda_\Delta v_\Delta v_\varphi + \mu_H^2}{\lambda_H}\right)^{1/2}\,, \quad v_\Delta \simeq \frac{\lambda_\Delta v_H^2 v_\varphi}{2 (M^2 + (\lambda_3 +\lambda_4)v_H^2/2)}\,, \quad v_\varphi \simeq \frac{\mu_\varphi}{\lambda_\varphi^{1/2}} \,.
\end{align}
In addition the mass matrices in terms  of $(h, \Delta_R,\varphi_R)$,  $(G^0, \Delta_I, \varphi_I)$  and $(G^+, \Delta^+)$  can be diagonalised into  CP-even or odd spectrum by  respective orthogonal  matrices.  Analogously  the  inert bosons  $(s,\eta)_{R/I}$ and $(s',\eta')_{R/I}$ are written  as:
\begin{align} 
 \eta =\left[
\begin{array}{c}
\eta^{+}\\
\frac{\eta_R + i\eta_I}{\sqrt2}
\end{array}\right]\,, \quad 
s  = \frac{s_R+is_I}{\sqrt2}\,;  \quad  \eta' =\left[
\begin{array}{c}
\eta^{\prime +}\\
\frac{\eta'_R + i\eta'_I}{\sqrt2}
\end{array}\right]\,, \quad
s' = \frac{s^{\prime}_R+is^{\prime}_I}{\sqrt2}\,.
\label{component}
\end{align}
They are rotated into the mass  basis  as follows:  
{\footnotesize{
\begin{align}
&V_\alpha^{T}\left[\begin{array}{cc}
m_{s_R}^2 &  \frac{ \lambda_0}{2} v_H v_\varphi  \\ 
\frac{ \lambda_0}{2}  v_H v_\varphi & m_{\eta_{R}}^2 \\ 
\end{array}\right] V_\alpha = \left[\begin{array}{cc} m^2_{H_1} & 0 \\ 0 & m^2_{H_2} \end{array}\right], ~~
V_{\alpha'}^{T} \left[\begin{array}{cc}
m_{s'_R}^2 &  \frac{\lambda'_0}{2}  v_H v_\varphi  \\ 
\frac{\lambda'_0}{2}  v_H v_\varphi & m_{\eta'_{R}}^2 \\ 
\end{array}\right] V_{\alpha'} = \left[\begin{array}{cc} m^2_{H'_1} & 0 \\ 0 & m^2_{H'_2} \end{array}\right] \\ \nn \\
&\left[\begin{array}{c} s_R + i s_I \\ \eta_R + i \eta_I \end{array}\right] = 
\left[\begin{array}{cc} c_{\alpha} & -s_{\alpha} \\ s_{\alpha} & c_{\alpha} \end{array}\right]
\left[\begin{array}{c} H_1 + i A_1 \\ H_2 + i A_2 \end{array}\right],\quad
\left[\begin{array}{c} s'_R + i s'_I \\ \eta'_R + i \eta'_I \end{array}\right] = 
\left[\begin{array}{cc} c_{\alpha'} & -s_{\alpha'} \\ s_{\alpha'} & c_{\alpha'} \end{array}\right]
\left[\begin{array}{c} H'_1 + i A'_1 \\ H'_2 + i A'_2 \end{array}\right]
\end{align}}}
where we use  shorthands of $s_{\alpha^{(')}} =  \sin\alpha^{(')}$, $c_{\alpha^{(')}} =  \cos\alpha^{(')}$ and the complex fields $H_{i} + i A_{i}$, $H^\prime_i + i  A^{\prime}_{i}$, $i = 1,2$ are mass eigenstates. Note that the semi-annihilation  exists for the  theory  with a   $\mathbb{Z}_{3}$  parity,  indicating that we need to keep  the degeneracy between  $H_{1,(2)}$ and $A_{1,(2)}$.  The reason is that  a $\mathbb{Z}_{3}$ parity assignment $w = e^{i 2 \pi /3}$ is valid for a Dirac fermion or a complex scalar field, like $\tilde{H} _i= H_i + i A_i$ with $i = 1,2$.  Under this specific potential we obtain that:   $m^2_{H_{1,2}}=m^2_{A_{1,2}}$, $m^2_{H'_{1,2}}=m^2_{A'_{1,2}}$. Without loss of generality, we can assume  $m_{H_1} < m_{H_2}$ and $m_{H'_1}<m_{H'_2}$ by ordering the mass eigenstates.

\begin{figure}[tb]
\begin{center}
\includegraphics[scale=0.48]{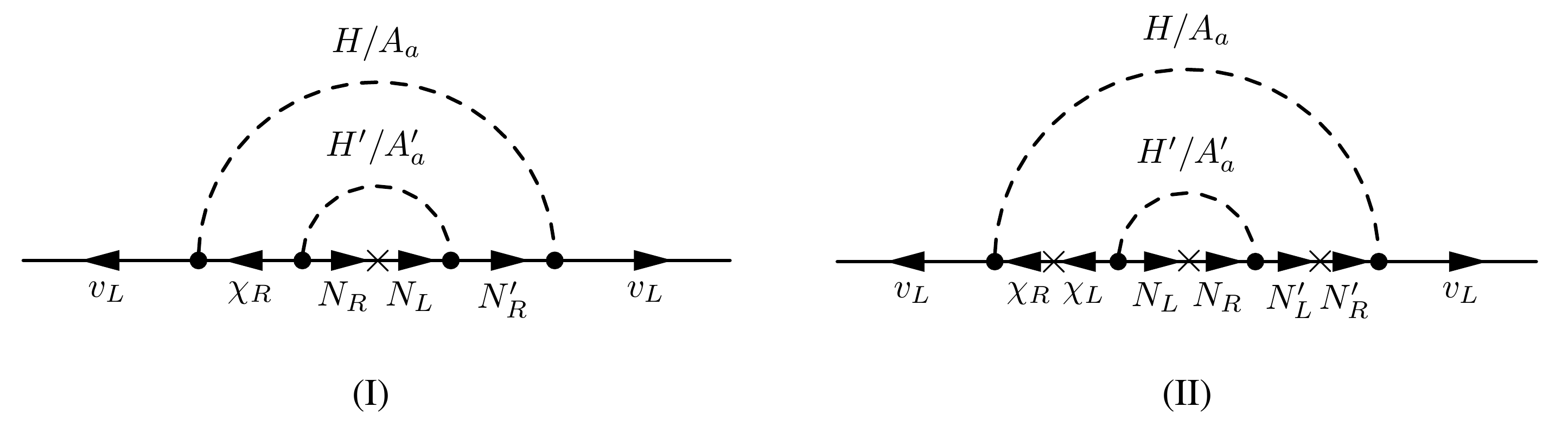}
\caption{ The Feynman diagrams for neutrino masses generated at the two-loop level in the mass eigenstate basis of inert scalars.
}   \label{fig:neutrino}
\end{center}\end{figure}
\subsection{ Neutrino mass matrix}
In this model, the neutrino mass arises  at the 2-loop level. 
To facilitate the calculation,  the  Lagrangian should be transformed  into the mass basis:
\begin{align}
- \mathcal{L}_{Y}&\sim
\frac{y_{\eta_{ia}}}{\sqrt2} \bar\nu_{L_i}\chi_{R_a}(s_{\alpha} H_1+c_{\alpha} H_2)
-i\frac{y_{\eta_{ia}}}{\sqrt2} \bar\nu_{L_i}\chi_{R_a}(s_{\alpha} A_1+c_{\alpha} A_2)\nn\\
&+
\frac{y_{\chi_{ab}}^{L/R}}{\sqrt2} \bar\chi_{L_a/R_a}^C\chi_{L_a/R_b}(c_{\alpha} H_1- s_{\alpha} H_2)
+i\frac{y_{\chi_{ab}}^{L/R}}{\sqrt2} \bar\chi_{L_a/R_a}^C\chi_{L_a/R_b}(c_{\alpha} A_1 - s_{\alpha} A_2)\nn\\
&+
\frac{y_{S_{ia}}}{\sqrt2} \bar\nu_{L_i} N'_{R_a}(c_{\alpha} H_1- s_{\alpha} H_2)
+i\frac{y_{S_{ia}}}{\sqrt2} \bar\nu_{L_i}N'_{R_a}(c_{\alpha} A_1 - s_{\alpha} A_2)\nn\\
&+
\frac{y_{s'_{ab}}}{\sqrt2} \bar N_{R_a} \chi^C_{R_b} (c_{\alpha'} H'_1- s_{\alpha'} H'_2)
+i\frac{y_{s'_{ab}}}{\sqrt2} \bar N_{R_a} \chi^C_{R_b} (c_{\alpha'} A'_1 - s_{\alpha'} A'_2)\nn\\
&+
\frac{y'_{s'_{ab}}}{\sqrt2} \bar N_{L_a} \chi^C_{L_b} (c_{\alpha'} H'_1- s_{\alpha'} H'_2)
+i\frac{y'_{s'_{ab}}}{\sqrt2} \bar N_{L_a} \chi^C_{L_b} (c_{\alpha'} A'_1 - s_{\alpha'} A'_2)\nn\\
%
&+
\frac{y_{\eta'_{ab}}}{\sqrt2} \bar N'_{R_a} N_{L_b} (s_{\alpha'} H'_1+ c_{\alpha'} H'_2)
-i\frac{y_{\eta'_{ab}}}{\sqrt2} \bar N'_{R_a} N_{L_b} (s_{\alpha'} A'_1 + c_{\alpha'} A'_2)\nn\\
&+
\frac{y'_{\eta'_{ab}}}{\sqrt2} \bar N'_{L_a} N_{R_b} (s_{\alpha'} H'_1+ c_{\alpha'} H'_2)
-i\frac{y'_{\eta'_{ab}}}{\sqrt2} \bar N'_{L_a} N_{R_b} (s_{\alpha'} A'_1 + c_{\alpha'} A'_2)+{\rm h.c.}.
\label{Eq:lag-mass}
\end{align}
Here we assume that all the Yukawa couplings are real for simplicity.  The active neutrino mass matrix $m_{\nu_{ij}}$  are generated at two-loop level as shown in Figure~\ref{fig:neutrino}, with their formulas given by 
\begin{align}
&(m_{\nu})_{ij}
=m_{\nu_{ij}}^{(I)} + m_{\nu_{ij}}^{(II)} +[m_{\nu_{ij}}^{(I)}]^T + [m_{\nu_{ij}}^{(II)} ]^T,
\end{align}
where $m_{\nu_{ab}}^{(I)}$ and $m_{\nu_{ab}}^{(II)} $ respectively correspond to the left and right plots in  Figure~\ref{fig:neutrino}. The constraint on the neutrino matrix  is  from the neutrino oscillation data,  since $({m}_\nu)_{ab}$ have to be diagonalized by the Pontecorvo-Maki-Nakagawa-Sakata mixing matrix $V_{\rm MNS}$ (PMNS)~\cite{Maki:1962mu} as $({m}_\nu)_{ij} =(V_{\rm MNS}^* D_\nu V_{\rm MNS}^\dag)_{ij}$ with $D_\nu = \rm{diag} (m_{\nu_1}, m_{\nu_2}, m_{\nu_3})$. The PMNS matrix is parametrised as:
\begin{align}
V_{\rm MNS}&=
\left[\begin{array}{ccc} {c_{13}}c_{12} &c_{13}s_{12} & s_{13} e^{-i\delta}\\
 -c_{23}s_{12}-s_{23}s_{13}c_{12}e^{i\delta} & c_{23}c_{12}-s_{23}s_{13}s_{12}e^{i\delta} & s_{23}c_{13}\\
  s_{23}s_{12}-c_{23}s_{13}c_{12}e^{i\delta} & -s_{23}c_{12}-c_{23}s_{13}s_{12}e^{i\delta} & c_{23}c_{13}\\
  \end{array}
\right] \nn \\
& \times \rm{diag}(1, e^{i \frac{\alpha_{21}}{2}} , e^{i \frac{\alpha_{31}}{2}} )
\end{align}
with $s_{ij} = \sin \theta_{ij}$ being three mixing angles. In the following analysis,  we will  also neglect the Majorana CP violation phase  $\alpha_{21}$ and $\alpha_{31}$ as well as Dirac CP violation phase $\delta$. By assuming the  normal mass order $m_{\nu 1} \ll m_{\nu 2} < m_{\nu 3}$, the global fit of the current experiments at $3 \sigma$ is given by~\cite{pdg2018}:
\begin{eqnarray}
&& 0.250 \leq s_{12}^2 \leq 0.354, \; 
 0.381 \leq s_{23}^2 \leq 0.615, \;
 0.019 \leq s_{13}^2 \leq 0.024,  \nn \\
&& 
 m_{\nu_3}^2- m_{\nu_1}^2 =(2.45 - 2.69) \times10^{-3} \ {\rm eV}^2,  \nn  \\
&&  \ m_{\nu_2}^2- m_{\nu_1}^2 = (6.93 - 7.96) \times10^{-5} \ {\rm eV}^2, 
  \end{eqnarray}
Now we rewrite the neutrino mass matrix in terms of Yukawa couplings and the form factors:
\begin{align}
&(m_{\nu})_{ij}\equiv 
\frac{1}{(4\pi)^4}\left(y_{\eta_{ia}} [F_I + F_{II}]_{ab} y^T_{S_{bj}} +  y_{S_{ja}} [F_I^T + F_{II}^T]_{ab} y^T_{\eta_{bj}}\right)\nn\\
&\equiv\frac{1}{(4\pi)^4}\left(y_{\eta_{ia}} G_{ab} y^T_{S_{bj}} +  y_{S_{ja}} G^T_{ab} y^T_{\eta_{bj}}\right), \label{eq:mnv}
\end{align}
where the factor $\frac{1}{(4 \pi)^4}$ comes from the loop integration and the exact expressions for these form factors  $F_I$, $F_{II}$  are put in  Appendix~\ref{app:neutrino}.  The form factors exhibit an  interesting property,  proportional  to the product of mass differences  $(m_{H_2}^2-m_{H_1}^2) (m_{H'_2}^{  2}-m_{H'_1}^{ 2})$. Thus the neutrino mass can be easily  accommodated into the sub-eV order,  if  either one set of inert scalars is quasi-degenerate without tuning the Yukawa couplings. In particular,  if we set $m_{H'_1} \simeq m_{H'_2}$,  the LFV bound will not be influenced  as $H'_{1,2}$ do not mediate these processes.

Due to the symmetric property, the Eq. (\ref{eq:mnv}) can be conveniently  recasted into a suitable form for the numerical analysis:
\begin{align}
y_{\eta}=\frac12[(V_{\rm MNS}^* D_\nu V_{\rm MNS}^\dag + A] (y^T_S)^{-1} G^{-1}, \label{eq:yeta}
\end{align}
where  the  $A$  is  an arbitrary anti-symmetric matrix in the order   $ \lesssim 10^{-9}$  and of complex values if there is CP violation~\cite{Okada:2015vwh}.  Therefore after we impose Eq.(\ref{eq:yeta}),  the $y_{\eta}$ coupling is no longer a free parameter but as a function of  $y_S$  and the neutrino mass form factors.  This parameter will be determined by  the neutrino oscillation data up to an uncertainty.  Notice  that $y_{\eta}\lesssim\sqrt{4\pi}$ should be satisfied in the perturbative limit.

\subsection{LFV  and Muon $g-2$}
\label{lfv-lu}
\begin{figure}[tb]
\begin{center}
\includegraphics[scale=0.48]{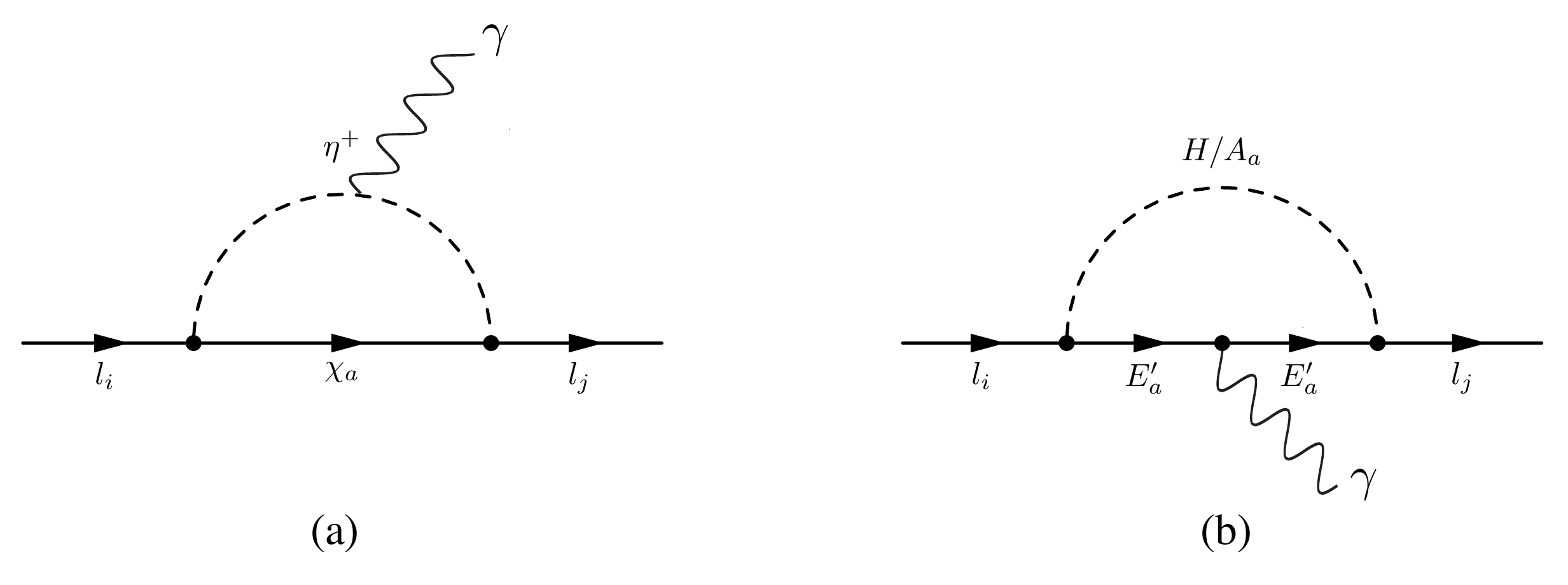}
\caption{ Lepton flavor violation processes induced by  heavy fermions and scalars.
}   \label{fig:flv}
\end{center}\end{figure}

In this radiative neutrino mass model, the existence  of charged scalars and vector-like fermions  contribute to lepton flavor violation processes  (see Figure.~\ref{fig:flv}), which in turn will severely constrain the Yukawa couplings and masses of heavy scalars and fermions. The relevant Lagrangian for LFV  can be expressed as:
\begin{align}
{\cal L}={-} y_{\eta_{ia}}\bar\ell_{L_i} \eta^- \chi_{R_a} + \frac{1}{\sqrt2}y_{S_{ia}} \bar\ell_{L_i} E'_{R_a}
[(c_{\alpha_R} H_1 - s_{\alpha_R} H_2) +i(c_{\alpha_I} A_1 - s_{\alpha_I} A_2)]+{\rm h.c.},
\end{align}
We can calculate  the branching ratio for LFV decay process $\ell_i \to \ell_j \gamma$ in terms of amplitude  $a_{L/R}$, which encodes the loop integration of the Feynman diagrams:
\begin{align}
Br(\ell_i\to\ell_j \gamma)
\approx 
\frac{48\pi^3 \alpha_{\rm em}}{{G_{\rm F}^2 m_{\ell_i}^2} } C_{ij}\left( |a_{L_{ij}}|^2 + |a_{R_{ij}}|^2\right),
\end{align}
where $G_{\rm F}\approx1.166\times 10^{-5}$ GeV$^{-2}$ is the Fermi constant, 
$\alpha_{\rm em}(m_Z)\approx {1/128.9}$ is the 
fine-structure constant~\cite{pdg2018},  $C_{21}\approx1$, $C_{31}\approx0.1784$, and $C_{32}\approx0.1736$.
In this specific model $a_R$ is formulated as:
\begin{align}
& a_{R_{ij}}
\approx   \frac{m_{\ell_i}}{(4\pi)^2} 
\left[ y_{\eta_{ja}} y^\dag_{\eta_{ai}} H(\chi_a,\eta^-) \right.
\nn \\ & \left. -\frac{y_{S_{ja}} y^\dag_{S_{ai}}}{2}
[c^2_{\alpha_R} H(H_1,E'_a) + s^2_{\alpha_R} H(H_2,E'_a)+c^2_{\alpha_I} H(A_1,E'_a)+s^2_{\alpha_I} H(A_2,E'_a)]
\right], \label{eq:lfv}\\
&H(a,b)=\int_0^1dx\int_0^{1-x}dy\frac{xy}{x ~m^2_a +(1-x) ~ m^2_b} \nn \\
& \quad \quad  \overset{m_a \neq m_b}{=} \frac{2 m_{a}^6+3 m_{a}^4 m_{b}^2-6 m_{a}^2 m_{b}^4+ m_{b}^6 + 6 m_{a}^4 m_{b}^2 \log
   \left(\frac{m_{b}^2}{m_{a}^2}\right)}{12 \left(m_{a}^2- m_{b}^2\right)^4} 
\label{eq:lfv-lp}
\end{align} 
where we can see that  the loop contributions from two resources (Figure 2.a and Figure 2.b) are in  opposite signs.  And for the left-handed amplitude,   $a_L$  is  obtained  by a mass substitution: $a_{L}=a_{R}(m_{\ell_i}\to m_{\ell_j})$.

The couplings involved in those LFV processes are $y_\eta$ and $y_S$,  strongly correlated to the neutrino mass matrix. In particular the  magnitude of $y_\eta$ along with  masses   $m_{\chi_1}$ and $m_{H_{1,2}}$, constrained by  the LFV  bound,  will  influence the DM relic density as well.  To find out  the allowed parameter space for this  model, the following upper bounds are imposed~\cite{TheMEG:2016wtm, Aubert:2009ag}
  \begin{align}
 & \quad \quad  \quad Br(\mu\rightarrow e\gamma) \leq4.2\times10^{-13} ~ (6\times10^{-14})   \nn \\
   & Br(\tau\rightarrow \mu\gamma)\leq4.4\times10^{-8}, \quad
  Br(\tau\rightarrow e\gamma) \leq3.3\times10^{-8} 
 \label{expLFV}
 \end{align}
where the upper bound from $\mu \to e  \gamma $ is the most stringent one with the value in  parentheses  indicating a future reach of MEG experiment~\cite{Renga:2018fpd}. 

{\it The muon anomalous magnetic moment}:  The muon $g-2$ is a well-measured property and a large $ 3.6 \sigma$ discrepancy of $\Delta a_\mu$  between the SM theory and experiment measurement was observed for a long time. For this model, one can  estimate the muon $g-2$  through the amplitudes formulated above: 
\begin{align}
\Delta a_\mu  \approx -m_\mu (a_L+a_R)_{22}.\label{eq:damu}
\end{align}
The  deviation from the SM prediction is  
$\Delta a_\mu = a_\mu^{\rm exp} - a_{\mu}^{ \rm SM} =(2.74\pm 0.73)\times10^{-9}$~\cite{pdg2018} with a positive value.
However because  our  analysis shows  the muon $g-2$ is too tiny after imposing  other bounds,  we just  employ the muon $g-2$  as a model quality for reference. 

\subsection{$Z-Z'$ mixing}\label{sec:zzH}

The effect of the hidden $Z'$  at TeV scale  will actually decouple from the dark matter physics and we would like to qualify this argument in the section. After the three scalar fields developing VEVs, $U(1)_H$ and electroweak symmetries are spontaneously broken so that the mass terms of neutral gauge boson are obtained,
\begin{align}
 \frac{1}{ 2 }\begin{pmatrix}  Z_0 \\ \tilde Z \end{pmatrix}^T 
\left[\begin{array}{cc}
\frac{(g_1^2+g_2^2)}{4} (v_H^2+4 v_{\Delta}^2) &  3 x \sqrt{g_1^2+g_2^2}g_H v_{\Delta}^2  \\ 
3 x \sqrt{g_1^2+g_2^2}g_H v_{\Delta}^2  & 9 x^2 g_H^2 (v_{\Delta}^2+  v_{\varphi}^2)   \\ 
\end{array}\right] \begin{pmatrix}  Z_0 \\ \tilde Z \end{pmatrix},
\end{align}
where $g_2$, $g_1$ and $g_H$ are gauge couplings of  $SU(2)_L$, $U(1)_Y$, and $U(1)_H$, respectively. The 
$Z_0 $ and $\tilde Z$ are the gauge fields for $U(1)_Y$ and  $U(1)_H$ with the $Z_0$ mostly composed of the SM Z boson. Here we  assume the  kinetic mixing between the two Abelian gauge bosons to be negligibly small for simplicity.
In case of  $x =1$, we parameterise the mass matrix to be:
\begin{equation}
\left[\begin{array}{cc}
\frac{(g_1^2+g_2^2)}{4} (v_H^2+4 v_{\Delta}^2) &  3 ~ \sqrt{g_1^2+g_2^2}g_H v_{\Delta}^2  \\ 
3 ~ \sqrt{g_1^2+g_2^2}g_H v_{\Delta}^2  & 9 ~ g_H^2 (v_{\Delta}^2+  v_{\varphi}^2)   \\ 
\end{array}\right] 
=
m_{\tilde Z}^2
\left[\begin{array}{cc}
\epsilon_1^2 & 2 \epsilon_1 \epsilon_2 \epsilon_3  \\ 
2 \epsilon_1 \epsilon_2 \epsilon_3 & 1+\epsilon_2^2  \\ 
\end{array}\right],
\end{equation}
where we use the definition of  $m_{Z_{0}} =  \frac{\sqrt{(g_1^2+g_2^2)\left(v_H^2+ 4 v_\Delta^2\right)}}{2}$, $m_{\tilde Z} =  3 g_H v_\varphi $, $\epsilon_1 = \frac{m_{Z_{0}}}{m_{\tilde Z}}$ and $\epsilon_2 =  \frac{v_{\Delta}}{v_\varphi}$, $\epsilon_3 = \frac{v_\Delta}{\sqrt{v_{H}^2 + 4 v_{\Delta}^2}}$.
The mass matrix can be diagonalized by an orthogonal transformation to  be ${\rm Diag}(m^2_{Z},m^2_{Z'}) $,
and in an approximation of  $v_\Delta \ll v_{H} \lesssim  v_{\varphi}$ and $g_H = \mathcal{O}(1)$, this gives:
\begin{align}
 m^2_{Z}  & \approx m_{Z_{0}}^2 \left(1-  4 \epsilon_2^2 \epsilon_3^2   \right),\,  \quad
m^2_{Z'}\approx m_{\tilde Z}^2 \left(1 +  \epsilon_2^2  \right),\label{eq:zm}
\\
\begin{pmatrix}  Z \\  Z' \end{pmatrix}  & = 
\left[\begin{array}{cc}
c_{Z} &  s_{Z} \\ 
-s_{Z}  &  c_{Z}  \\ 
\end{array}\right] \begin{pmatrix}  Z_0 \\  \tilde{Z} \end{pmatrix} , \quad \tan \theta_{Z} = \frac{-2 \epsilon_1 \epsilon_2 \epsilon_3}{1+\epsilon_2^2-\epsilon_1^2} .\label{eq:tan}
\end{align} 
If we fix   $c_W^2 =  g_2^2/(g_1^2+g_2^2) $ as the SM value,  the $\rho$ parameter  can be expressed to be:
\begin{align}
\rho_0 \simeq \frac{\left(1+ \frac{2 v_{\Delta}^2}{v_{H}^2} \right)}{ \left(1+ \frac{4 v_{\Delta}^2}{v_{H}^2} \right) \left( 1- 4 \epsilon_2^2 \epsilon_3^2 \right)}
\end{align}
The experimental constraint from the PDG is  $\rho_{0,\, \rm{exp}} = 1.00039 \pm 0.00019$~\cite{pdg2018},  which will translate into a requirement of $v_{\Delta} \lesssim 3.5 $ GeV. In this paper,  we assume the $Z'$ boson mass to be above the TeV scale for  $v_\varphi  \gtrsim 350$ GeV.  According to Eq.~(\ref{eq:tan}), this results in  a extremely small  $|\tan \theta_Z| < 10^{-5}$ compared with the Yukawa coupling with DM and neutrino. Thus as long as we prefer  the DM mass in  $\mathcal{O}(100)$ GeV,  it will be safe to neglect the the  impact of $Z'$ on either DM annihilation or  DM-nucleon scattering,

\subsection{ Bound of Electroweak Precision Test}

The Electroweak Precision Test (EWPT) on low energy observables can set limits for deviations from the SM. The new physics effects are mainly encoded the  oblique parameters $S$, $T$ and $U$, expressed  in terms of the transverse part of  gauge boson's self-energy amplitudes. For this model, since the $U$ parameter is suppressed by an additional  factor of $v^2/M_{\mbox{new}}^2$,  its effect is neglected. Due to the vector-like nature and  degeneracy, the exotic particles of $L'_i$ have no impact on oblique parameters, i.e. $\Delta S_f = \frac{2}{3 \pi} (t_{3L}-t_{3R})^2 = 0$ and $\Delta T_f =0$~\cite{pdg2018}.   However the inert scalars $(\eta, s)$ are possible to  cause notable deviation to   $S = -16 \pi \Pi^\prime (0)_{W_3 B}$ and $T = \frac{4 \pi}{m_Z^2 s^2_W c^2_W } \left[ 2 \Pi_{W_1 W_1}(0) - \Pi_{W_3 W_3} (0)\right]$~\cite{Peskin:1991sw},  we will  discuss  their constraints on the mass splitting among $(m_{\eta^+} ,m_{H_i})$ and the mixing angle  $\sin (\alpha)$.  After  evaluating the relevant self-energy correlations, we find out  the effects from the inert scalars are described by:
\begin{eqnarray}
\Delta S  &  = & \frac{1}{1 2\pi }\left[s_{\alpha}^4 G(m_{H_1},m_{H_1}, m_{\eta^+})+ c_{\alpha}^4 G(m_{H_2}
m_{H_2}, m_{\eta^+}) + 2 c_{\alpha}^2 s_{\alpha}^2 G(m_{H_1},m_{H_2}, m_{\eta^+})\right] \nn \\
&=& \frac{1}{12 \pi} \left[s_{\alpha}^2 \ln \left(\frac{m_{H_1}^2}{m_{\eta^+}^2}\right)+ c_{\alpha}^2  \ln \left(\frac{m_{H_2}^2}{m_{\eta^+}^2} \right) - 3 c_{\alpha}^2 s_{\alpha}^2 \chi (m_{H_1}, m_{H_2})\right] \\
\Delta T  &  = & \frac{1}{16 \pi m_W^2 s_W^{2}}[  s_{\alpha}^2 F (m_{H_1},m_{\eta^+})+  c_{\alpha}^2 F (m_{H_2}
m_{\eta^+}) -   c_{\alpha}^2 s_{\alpha}^2 F(m_{H_1},m_{H_2})] \label{eq:ST}
\end{eqnarray}
where the loop functions $G(m_{1}, m_{2}, m_{3})$, $\chi(m_1,m_2)$ and $F(m_1, m_2)$ are defined as:
\begin{align}
& G(m_{1}, m_{2}, m_{3}) =   \frac{1}{2} \left[\ln \left(\frac{m_1^2 m_2^2}{m_3^4}\right)- 3 \chi (m_1, m_2)\right] \\
&  \chi  (m_1, m_2) =   \frac{5 \left(m_1^4+m_2^4\right)-22 m_1^2 m_2^2}{9 \left(m_1^2- m_2^2\right)^2}  + \frac{3 m_1^2 m_2^2 \left(m_1^2+m_2^2\right) -m_1^6 - m_2^6  }{3 \left(m_1^2- m_2^2\right)^3} \ln  \left(\frac{m_1^2}{m_2^2}\right) \\
&  F(m_{1},m_{2})=m_{1}^{2}+m_{2}^{2}  -\frac{2 m_{1}^{2}m_{2}^{2}
}{m_{1}^{2}-m_{2}^{2}}\ln \left(\frac{m_{1}^{2}}{m_{2}^{2}} \right).
\end{align}
with  $\chi (m_1, m_2) $ and $F(m_1, m_2)$ being symmetric for $m_1 \leftrightarrow m_2 $ and vanishing for equal masses, i.e. $\chi(m,m) = F(m,m) = 0 $. During the calculation, the divergences inherent in the two-point functions are properly cancelled~\footnote{For the $T$ parameter, if we calculate it in terms of  the gauge boson's self-energy amplitudes,  the divergence is  fully captured in the loop function $A_0(m^2) = \frac{1}{i \pi^2}\int  \frac{d^4 k}{(k^2- m^2)}$~\cite{Haber:2010bw} and should be cancelled after counting all the diagrams. The cancellation due to the mixing neutral inert scalars $((\eta_R + i \eta_I)/\sqrt{2}), s)$  (precisely speaking, $(H_{1,2}, A_{1,2})$ in the mass basis) demonstrates in the following  pattern: $2 s^2_{\alpha} A_0(m_{H_1}^2)+2 c^2_{\alpha} A_0(m_{H_2}^2) - 2 s^4_{\alpha} A_0(m_{H_1}^2) - 2 c^4_{\alpha} A_0(m_{H_2}^2) - 2 c^2_\alpha s^2_\alpha (A_0(m_{H_1}^2)+A_0(m_{H_2}^2)) =0$.}.  In case of  the  SM Higgs $h$ barely mixing with $(\Delta, \varphi)$,  the $\Delta T$ is exactly  the wave-function renormalisation of  the goldstone bosons $G^+, G^0$ with $(\eta^\pm, ((\eta_R + i \eta_I)/\sqrt{2}), s)$ running inside the loops(referring to Appendix~\ref{app:Tpara} for detail) ~\cite{Barbieri:2006dq}. While for the  $\Delta S$, the function $G(m_1,m_2, m_3)$ is related to $\frac{d}{d p^2} \left [ B_{22} (p^2, m_1^2, m_2^2) -  B_{22} (p^2, m_3^2, m_3^2)\right] |_{p^2=0}= \frac{1}{2} \int_0^1 dx ~ x(1-x) \ln [\frac{x m_1^2 + (1-x) m_2^2}{m_3^2}]$, using the Passarino-Veltman function $B_{22}$  defined in~\cite{Passarino:1978jh}. 

\begin{figure}[tb]
\begin{center}
\includegraphics[height=5.5 cm,  width=6.6cm]{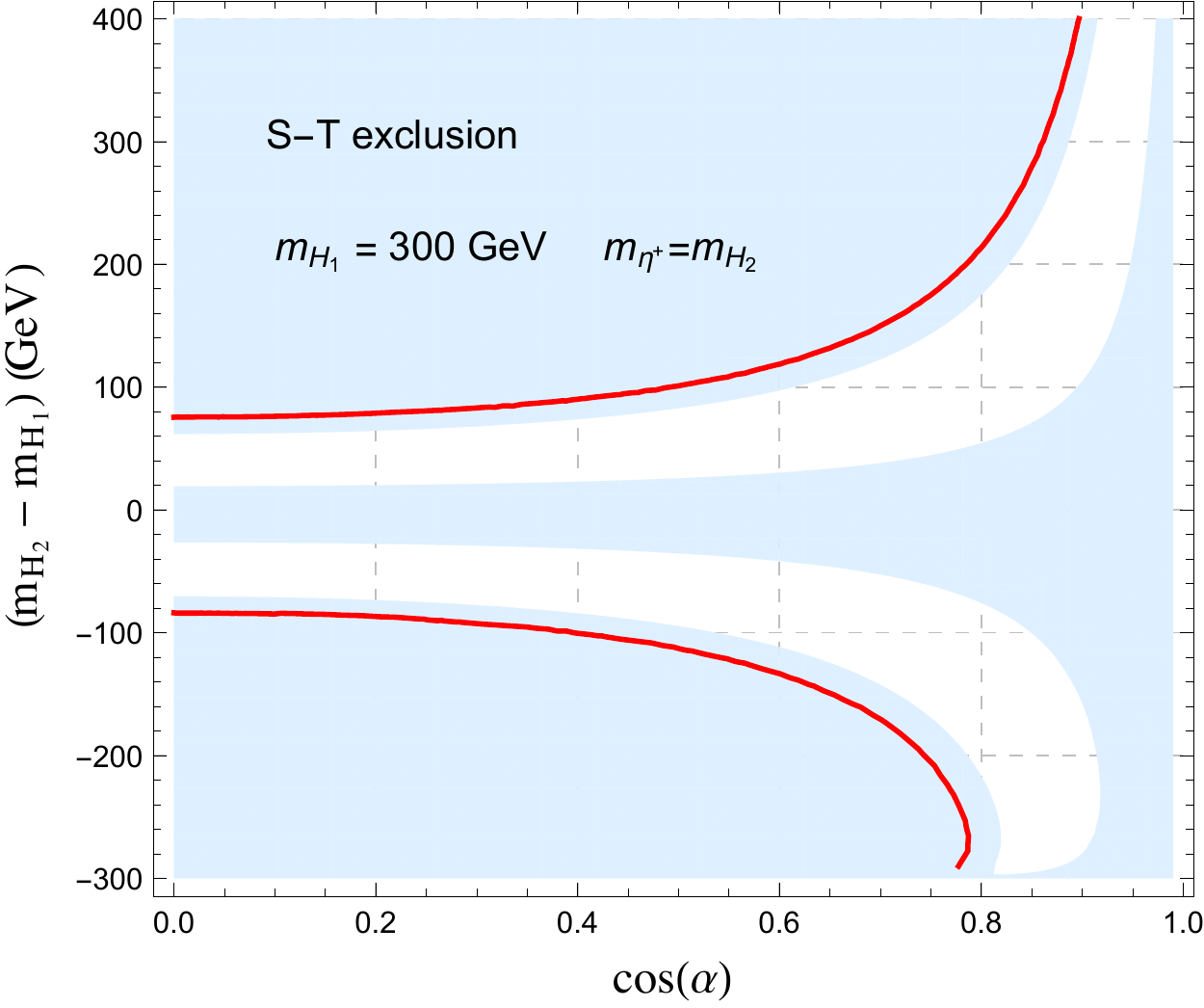} \quad \quad
\includegraphics[scale=0.5]{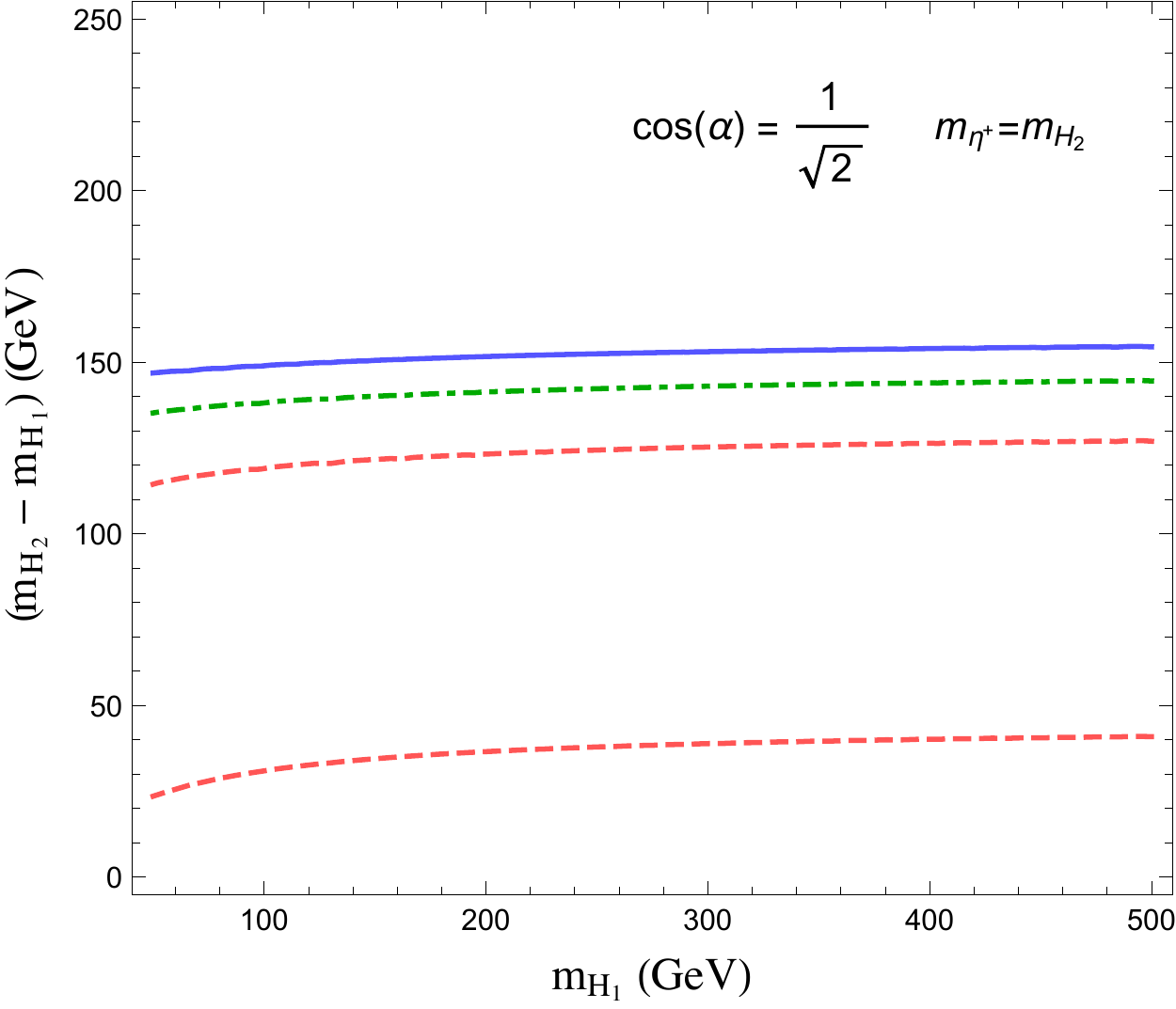}
\vspace{-0.18 cm}
\caption{In the left plot, the two white bands present  the region  allowed  by the $S-T$ bound at the $1 \sigma$ C.L.;  While  the red line is the $3 \sigma$ bound  for $m_{H_2} = 300$ GeV. The right plot shows the bound for $(m_{H_2}- m_{H_1})$ and  $m_{H_1}$ in case of $s_\alpha = c_\alpha = \frac{1}{\sqrt{2}}$ (assuming $m_{H_2} > m_{H_1} $ and $m_{\eta^+} = m_{H_2}$), where the regions outside the contours of  red, green and blue lines are excluded at $68 \% (1 \sigma) $,  $95\% (2 \sigma)$ and $99\% (3 \sigma)$ C.L.s. }
\label{fig:ST}
\end{center}
\end{figure}

The bound for the $S$ and $T$ parameters is obtained from the precision electroweak data, such as $M_{Z}$ and $\Gamma_Z$, at the $1 \sigma$ deviation by fixing $U=0$~\cite{pdg2018}:
\begin{align}
S = 0.02 \pm 0.07 \qquad  T = 0.06 \pm 0.06
\end{align}
with an off-diagonal  correlation of $92\%$. In Figure~\ref{fig:ST}, combining all the discussed parts,  we translate the  $S$-$T$   $\chi^2$  fit into  the bounds of  inert scalar masses  and the mixing angle. Since both $\Delta S$ and $\Delta T$ are symmetric functions of $(m_{H_1} , m_{H_2})$, we focus on the case of $m_{\eta^+} = m_{H_2}$ for  simplicity. The operation of  switching $H_1 \leftrightarrow H_2 $ is   to shift the mixing angle by $\alpha \to (\alpha -\frac{\pi}{4})$. the $1 \sigma$ EWPT fit  prefers the mass splitting  $(m_{H_2} - m_{H_1})>0$ in a small range of $0.92 < \cos \alpha <1.0$ , i.e. $H_2 (A_2)$ dominantly composed of $\eta_R (\eta_I) $ should be heavier.  However at $3 \sigma$ fit,   $(m_{H_2} - m_{H_1})$ is permitted in either sign $(+/-)$ for  $0 < \alpha<\frac{\pi}{2} $, with its  magnitude  decreasing with $\cos \alpha$.  In the right plot, we show that assuming $m_{H_2} > m_{H_1}$, the  $S$-$T$ bound  requires  $(m_{H_2} - m_{H_1}) \subset (30, 120)$ GeV at  $1 \sigma$ and $(m_{H_2} - m_{H_1})< 150$ GeV at $3 \sigma$  for $100 < m_{H_1} < 500$ GeV under the condition specified in the caption.

\section{Dark matter }

The  relic density for a DM specie $X$  is determined by  its  energy density, $\propto  m_X n_X(T_0) $ in the present universe, where the number density $n_X$ is   governed by the Boltzmann equation during the decoupling phase  plus the afterwards expansion effect. For a Dirac fermion DM stabilised by a $\mathbb{Z}_3$ symmetry,   semi-annihilation modes in addition to annihilation are expected to contribute.  The Boltzmann equation can be recasted into an evolution  in terms of  a yield   by defining  $Y_X = n_X / s$ with  $s$  to be entropy density and $x = M_X/T$ where the temperature is scaled by the DM mass. The redefined equation reads:
\begin{eqnarray}
&& \frac{d Y_X}{d x}  = -\frac{\lambda_{\rm A}}{x^2}[Y_X^2 - Y_X^{\rm{eq} 2}]  - \frac{1}{2} \frac{\lambda_{\rm S}}{x^2}  [Y_X^2 - Y_X Y_X^{\rm{eq} }] \,, \label{eq:Boltzman} \\
&&  \lambda_i = \frac{s(x=1)}{H(x=1)} \langle \sigma v_{\rm rel} \rangle_i, \quad i = A, S \nn \\
&&  s(x = 1) = \frac{2 \pi^2}{45} g_* M_X^3  \,, \quad H (x = 1) = \sqrt{\frac{\pi^2}{90} g_*} \frac{M_X^2}{M_{\rm pl}} 
\end{eqnarray}
where $A,S$ stand for  annihilation and   semi-annihilation,  $H(x=1)$ is the Hubble constant at $T = M_X$,  $g_*$ is the effective total number of relativistic degrees of freedom and $M_{\rm pl}=1.22\times 10^{19}[{\rm GeV}] $ is the Planck mass.  The $\frac{1}{2}$ factor in the second term of Eq.~(\ref{eq:Boltzman}) is due to  the  identical {\it initial} particles~\footnote{For the semi-annihilation,  considering  the evolution of number density for one specie $X$, we need  take into account  the processes of  $X X \to \bar X \nu_i$ and $\bar X \bar X \to  X \bar{\nu}_i$,  where the number of the specie $X$  is  only depleted by ~$1$ in the  forward   direction,   same as  in the  particle-antiparticle annihilation. Thus the Boltzmann equation with only semi-annihilation mode  should be: $\frac{d n_X}{d t} + 3 H n_{X} =  - \frac{1}{2} \langle \sigma v \rangle_{\rm Semi}  [n_X^2 - n_X n_X^{\rm{eq} }]$.  This is different from the  DM annihilation of Majorana fermions, where the depletion number is $2$,  and compensates the phase space factor $\frac{1}{2}$ from identical particles.} and $\langle \sigma v_{\rm rel} \rangle$ is the thermal average of velocity weighted cross section which represents the DM interaction rate.  This equation can be analytically  solved in a proper approximation by  matching  the results from two regions at the freeze-out point. A brief review for this approach will be presented here in order to clarify the missing $1/2$ in some literature. We will start by defining  a quality $\Delta = Y_X - Y_X^{\rm eq}$, so that the original equation is transformed into:
\begin{eqnarray}
 \frac{d \Delta}{d x}  = -\frac{dY_X^{\rm eq}}{x} - \frac{\lambda_{\rm A}}{x^2}[\Delta^2 + 2 \Delta Y_X^{\rm{eq} }] - \frac{1}{2} \frac{\lambda_{\rm S}}{x^2}  [\Delta^2 +\Delta Y_X^{\rm{eq} }]  
\end{eqnarray}
where  the Maxwell-Boltzmann distribution will be used for the yield in equilibrium so that $Y_X^{\rm eq} (x) \propto x^{3/2} e^{-x}$. For $x \ll x_f$, we can obtain:
\begin{eqnarray}
\Delta =  \frac{Y_X^{\rm eq}}{\frac{\lambda_A}{x^2} (2 Y_X^{\rm eq} +\Delta) + \frac{\lambda_S}{2 x^2} (Y_X^{\rm eq} + \Delta)}
\end{eqnarray}
and for  $x \gg x_f$, the integration of Boltzmann equation gives:
\begin{eqnarray}
Y_X (\infty) \simeq - \int_{x_f}^\infty dx \frac{\lambda_A  + \frac{1}{2} \lambda_S }{x^2}
\end{eqnarray}
Thus the  relic density at the present universe  is found as:
\begin{align}
&\Omega h^2 = m_X s_0 Y_X (\infty) / \rho_c \approx 2\frac{1.07\times 10^9 {\rm GeV}^{-1}}{\sqrt{g_*(x_f)} M_{pl} J (x_f)},  \\
&J (x_f)= \int_{x_f}^\infty dx \frac{\langle\sigma v_{\rm rel}\rangle_{\rm A}  + \frac{1}{2} \langle\sigma v_{\rm rel}\rangle_{\rm S} }{x^2}  ,
\end{align}
where  $\Omega h^2$  is rescaled by the critical density $\rho_c = 3 H^2/8 \pi G$.  We times a  factor $2$ for the relic density in order to count  the contribution from the antiparticle $\bar{X}$ and set  $g_*(x_f)\approx 100$  at the point of freeze-out. Here $\langle\sigma v \rangle_{\rm A}$ is the thermal average for  annihilation, while $\langle\sigma v\rangle_{\rm S}$ is  for  {\it semi}-annihilation. Then the freeze-out  temperature $x_f$ is  determined by the boundary condition $\Delta (x_f) = c ~Y_X^{\rm eq} (x_f)$ with $c = \sqrt{2}-1$ to be:
\begin{align}
x_f\simeq \ln\left[0.038c(c+2)\langle\sigma v\rangle_{\rm A}  \frac{g M_X M_{pl}}{\sqrt{g_*x_f}}\right]
+\ln\left[1+\frac{c+1}{c+2}~\frac{\langle\sigma v\rangle_{\rm S}}{2 ~\langle\sigma v\rangle_{\rm A}}\right],\label{eq:xf}
\end{align}
which is up to a $1/2$ factor for  the semi-annihilation part as given by~\cite{DEramo:2010keq}  and we set  $g=2$ for a fermion DM of two degrees of freedom  without counting its antiparticle~\cite{Kolb}.

\begin{figure}[tb]
\begin{center}
\includegraphics[scale=0.8]{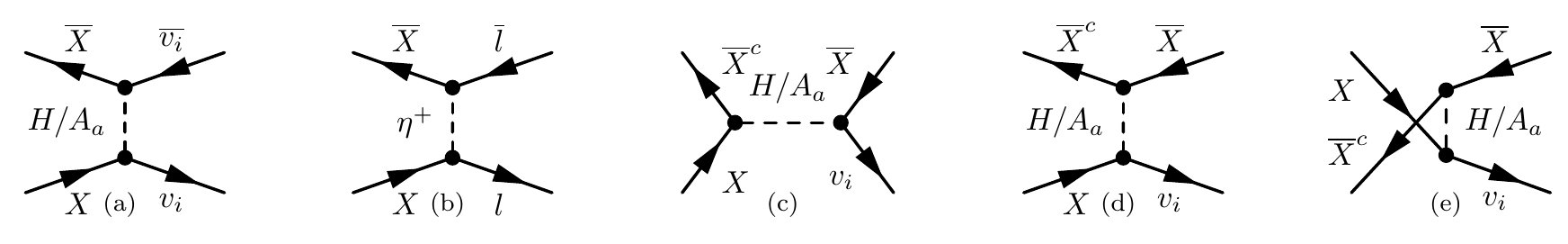}
\caption{Feynman diagrams for the annihilation (a)-(b) and semi-annihilation processes (c)-(e), where the mediating  scalar fields are $H_a$ or $ A_a$, with $a=1,2$.  }
\label{semi_anni}
\end{center}
\end{figure}

As we can see that in order to estimate the relic density,  one has to calculate the  thermal average of cross section  times the relative velocity $\langle \sigma v_{rel} \rangle$.  Generally the thermal average is approximated by an expansion in order of $x^{-n}$ ($\langle v^2 \rangle \sim \frac{6}{x} $  in the non-relativistic limit). However in our case, the dominant  DM cross section proceeds through an $S$-channel  with one very narrow  resonance  $\Gamma_M/ M_X \ll v_{\rm rel}$ and one  wider resonance $\Gamma_M/ M_X \sim v_{\rm rel}$.  Also for a  $S$-channel interaction mediated by a scalar,  the s-wave is vanishing for the velocity averaged cross section, thus the expansion in terms of $v_{\rm rel}^2$ is complicated to handle  for  two resonances interfering with each other.  We prefer to use the integration approach to evaluate $\langle\sigma v_{\rm rel}\rangle $ which is given  by~\cite{Gondolo:1990dk, Edsjo:1997bg} 
\begin{align}
&\langle\sigma v_{\rm rel}\rangle_A =\sum_{i=1}^2 \frac{\int_{4M^2_X}^\infty ~ ds ~ \sigma_{XX}^i (s-4M_X^2) \sqrt{s} K_1\left(\frac{\sqrt s}{M_X}x\right)}{8 M_X^5x^{-1}K_2(x)^2}  
\\
& \langle\sigma v_{\rm rel}\rangle_S = \frac{\int_{4M^2_X}^\infty ~ ds ~ \sigma_{XX}^3 (s-4M_X^2) \sqrt{s} K_1\left(\frac{\sqrt s}{M_X}x\right)}{8 M_X^5x^{-1}K_2(x)^2}
\end{align}
where $s= (k_1+k_2)^2$ is a Mandelstam variable and $K_{1,2}$ are the modified Bessel functions of  order 1 and 2 respectively. 
\begin{align}
&\sigma_{XX}^i  = \frac{|\bf{k}_1|}{32 \pi^2 s \sqrt{s- 4 M_X^2} } \int d\Omega |\bar {\cal M}_i|^2,  i = 1,2,3; \label{eq:sigma} \\
& \mbox{\rm with} ~~ |{\bf{k}}_1| = \sqrt{\frac{s}{4}-m_{l/\nu}^2} ~~ i = 1, 2;   \quad |{\bf{k}}_1| = \frac{s-M_X^2}{2 \sqrt{s}} ~~ i =3. \nn
\end{align}
 Here $\sigma_{XX}^i$ is the cross section of the $2\to 2$ process (denoting $\bf{k}_1$ as  3-momentum of  the first out-going particle) and  with the amplitude squared   $|{\cal \bar M}_{1,2}|^2$ corresponding to  $X  \bar X  \to \nu_i  \bar\nu_j $ and $X  \bar X  \to \ell_i  \bar\ell_j $  in  Fig.~\ref{semi_anni}(a-b)  and the third  $| {\cal \bar M}_{3}|^2$ standing for   $X X  \to  \bar X  \nu_i $, i.e. the  semi-annihilation  as depicted in Fig.~\ref{semi_anni}(c)-(e). 

We derive the analytic expression for each amplitude squared present in  Eq.(\ref{eq:sigma}). Let us consider the case that only the lightest flavor of $\chi_i$ is the DM candidate. By defining  $X = \chi_{1}$ and assuming $y^L_\chi=y^R_\chi$,  the  DM-scalar interaction in this model is written as:
\begin{align}
-{\cal L}&=
\frac{y_{\eta_{i1}}}{\sqrt2}\bar\nu_i P_R X(s_{\alpha} H_1 + c_{\alpha} H_2)
-i\frac{y_{\eta_{i1}}}{\sqrt2}\bar\nu_i P_R X(s_{\alpha} A_1 + c_{\alpha} A_2)
{-} y_{\eta_{i1}}\bar\ell_i P_R X\eta^-
\nn\\
&+\frac{y_{\chi_{11}}}{\sqrt2}\bar X^C  X(c_{\alpha} H_1 - s_{\alpha} H_2)
{+i}\frac{y_{\chi_{11}}}{\sqrt2}\bar X^C  X(c_{\alpha} A_1 - s_{\alpha} A_2)+{\rm h.c.}\,,
\end{align}
For the annihilation processes,  $|{\cal \bar M}_{1,2}|$  are the usual amplitude squared  with the spin  averaged for the initial states and summed  for the final states. However   a special treatment is needed  for  $|{\cal \bar M}_{3}|$  because of  the  identical incoming particles.  As illustrated in Fig.~\ref{semi_anni}(c)-(e),  the semi-annihilation proceeds in  $S$, $T$ and $U$ channels after  counting  the  momentum  exchanging  for the  identical {\it initial} particles.  In particular,  there is a symmetry factor $2$ for the $S$-channel amplitude~\footnote{We need consider the momentum exchanging   for the identical {\it initial}  particles due to the phase space integration in  thermal average. For  semi-annihilation $X(p_1) X (p_2)+X(p_2) X (p_1) \to \bar X (k_1) v_i (k_2)$,  the $S$-channel amplitude  is proportional to $[\bar{u^c}(p_1) u (p_2) - \bar{u^c}(p_2)u (p_1)] [ \bar{v} (k_1) u(k_2)] = 2~[ \bar{v}(p_1)u(p_2) ]~[\bar{v} (k_1)u(k_2)]$, where we use the identities $u^c = C \bar{u}^T =  v$  and $\bar{v}(p_2) u(p_1) = u^T (p_1) C^{-1} C \bar{v}^T(p_2) = - \bar{v}(p_1) u(p_2)$, with $C= i \gamma^0 \gamma^2$ being the charge conjugate operator. This is similar to the  identical scalar  case $\phi \phi \to H_{a}$, the symmetry factor is normally encoded  in the vertex.}. Combining all channels, we can arrive the following  analytic  expressions:

\begin{align}
|{\cal\bar M}_1|^2&=  \sum_{i,j=1}^3{|y_{\eta_{i1}}y_{\eta_{1j}}^\dag|^2}
\left| \frac{s_\alpha^{2}}{M_X^2-m_{H_1}^2-2p_1\cdot k_1} + \frac{c_\alpha^{2}}{M_X^2-m_{H_2}^2-2p_1\cdot k_1} 
\right|^2  (p_1\cdot k_1)(p_2\cdot k_2)\,,
\\
|{\cal\bar M}_2|^2&=  \sum_{i,j=1}^3
\left|\frac{y_{\eta_{i1}}y^\dag_{\eta_{1j}}}{M_X^2-m_{\eta^\pm}^2-2p_1\cdot k_1}\right|^2  
(p_1\cdot k_1)(p_2\cdot k_2) \,,
\end{align}
\begin{align}
|{\cal\bar M}_3|^2&=   {(s_{\alpha} c_{\alpha})^2}
\sum_{i=1}^3{|y_{\chi_{11}}y_{\eta_{i1}}|^2} \left[ 8 ~ | \sum_{a=1}^2 (-1)^{a+1}S_{inv}^a|^2 (p_1\cdot p_2 - M_X^2)( k_1\cdot k_2) 
\right. \nn \\ &\left.  +2 ~ |\sum_{a=1}^2 (-1)^{a+1}T_{inv}^a|^2 (p_1\cdot k_1 + M_X^2)(p_2\cdot k_2)  + 2~ |\sum_{a=1}^2 (-1)^{a+1}U_{inv}^a|^2 (p_2\cdot k_1 + M_X^2) p_1\cdot k_2 \right. \nn \\ & \left. +2 \sum_{a=1}^2 (-1)^{a+1} S^{Re,a}_{inv} \sum_{a=1}^2 (-1)^{a+1}T_{inv}^a
[(p_1\cdot p_2)(k_1\cdot k_2) - (p_1\cdot k_2)(p_2\cdot k_1)+(p_1\cdot k_1)(p_2\cdot k_2) \right. \nn \\ 
&\left. +M_X^2(- p_1\cdot k_2 + p_2\cdot k_2-  k_1\cdot k_2)]
+ 2 \sum_{a=1}^2 (-1)^{a+1} S^{Re,a}_{inv} \sum_{a=1}^2 (-1)^{a+1} U_{inv}^a[(p_1\cdot p_2)(k_1\cdot k_2)\right.\nn\\
&\left.-(p_1\cdot k_1)(p_2\cdot k_2) + (p_1\cdot k_2)(p_2\cdot k_1)
+M_X^2(- p_2\cdot k_2+  p_1\cdot k_2 -  k_1\cdot k_2)]\right.\nn\\
&\left.
-  \sum_{a=1}^2 (-1)^{a+1} T_{inv}^a \sum_{a=1}^2 (-1)^{a+1} U_{inv}^a [(p_1\cdot k_1)(p_2\cdot k_2)-(p_1\cdot p_2)(k_1\cdot k_2)
+ (p_1\cdot k_2)(p_2\cdot k_1)\right.\nn\\
&\left.
+M_X^2( k_1\cdot k_2 + p_1\cdot k_2 + p_2\cdot k_2)]\right] \,.  \label{eq:M3} 
\end{align}
In  the  ${\cal M}_3$ amplitude of semi-annihilation, we  define  $S_{inv}^a = 1/(s-m_{a}^2+im_{a}\Gamma_{a})$, $T_{inv}^a = 1/(2 M_X^2-m_{a}^2-2p_1\cdot k_1)$, $U_{inv}^a = 1/(M_X^2 - m_{a}^2-2p_1\cdot k_2)$ and  the index  $a = 1, 2$ corresponds to $H_1( A_1)$, $H_2 (A_2)$ respectively. The inner products are given in Appendix~\ref{app:formula}.

For the $S$-channel amplitude, the widths of inert scalars $H_{1,2}(A_{1,2})$ enter into the Breit-Wigner propagator  $S_{inv}^a$, whose magnitude  near two  on-shell poles $m_{H_1} = 2 M_X$ or $m_{H_2} = 2 M_X$ is determined by the  $\Gamma_{H_1}$ or $\Gamma_{H_2}$. Under this consideration  we will only be interested in the parameter region  $m_{H_1} < m_{H_2} < {\rm min} (m_{E'_i},  m_{\chi_2}, m_{\chi_3}) $  to ensure   a narrow resonance. Therefore the decay widths of $\Gamma_{H_1}$ ($= \Gamma_{A_1}$) and $\Gamma_{H_2}$  ($= \Gamma_{A_2}$) are  formulated  as:
\begin{eqnarray}
&& \Gamma_{H_1} = \theta(m_{H_1} -2 M_X) \Gamma ( H_1 \to X X + \bar{X} \bar{X} ) +  \theta(m_{H_1} -  M_X) \Gamma (H_1 \to X \bar{v_i} + \bar{X} v_i)  \nn  \\
&& \Gamma ( H_1 \to X  X + \bar{X} \bar{X}) = |y_{\chi_{11}}|^2 c_\alpha^2 \frac{(m_{H_1}^2 - 4 M_X^2)^{3/2}}{ 4 \pi ~m_{H_1}^2}  \nn \\
&& \Gamma (H_1 \to X \bar{v_i} + \bar{X} v_i) =  \sum_i^3 |y_{\eta_{i1}} y^\dagger_{\eta_{1i}}| s_\alpha^2 \frac{(m_{H_1}^2-M_X^2)^2}{16 \pi ~ m_{H_1}^3} 
\end{eqnarray}
and for  $H_2$, one more decay channel $H_2 \to H_1 h_0$, with  a coupling vertex of $\frac{1}{2}\lambda_0 v_\varphi (c_\alpha^2 -s_\alpha^2) =  s_{\alpha} c_{\alpha} (c_\alpha^2 -s_\alpha^2) (m_{H_1}^2 - m_{H_2}^2)/v_H $ and $h_0$ being the SM Higgs boson, will be open if it is  permitted by kinematics. 
\begin{eqnarray}
&& \Gamma_{H_2} = \theta(m_{H_2} -2 M_X) \Gamma ( H_2 \to X X + \bar{X} \bar{X} ) +  \theta(m_{H_2} -  M_X) \Gamma (H_2 \to X \bar{v_i} + \bar{X} v_i)  \nn \\
&& \quad \quad +  ~ \theta(m_{H_2} -M_{H_1} -m_{h_0}) \Gamma (H_2 \to H_1 h_0) \nn \\
&& \Gamma ( H_2 \to X  X + \bar{X} \bar{X} ) =  |y_{\chi_{11}}|^2 s_\alpha^2 \frac{(m_{H_2}^2 - 4 M_X^2)^{3/2}}{4 \pi ~m_{H_2}^2} \nn \\
&& \Gamma (H_2 \to X \bar{v_i} + \bar{X} v_i) = \sum_i^3 |y_{\eta_{i1}} y^\dagger_{\eta_{1i}}|  c_\alpha^2 \frac{(m_{H_2}^2-M_X^2)^2}{16 \pi ~ m_{H_2}^3} \nn \\
&&\Gamma (H_2 \to H_1 h_0)  = s_\alpha^2 c_\alpha^2 (c_\alpha^2-s_\alpha^2)^2 \frac{(m_{H_2}^2 -m_{H_1}^2)^2}{16 \pi ~ v_H^2 m_{H_2}^3}[(m_{H_2}^2 - (m_{H_1} +m_{h_0})^2) \nn \\ && \qquad \qquad \qquad  \qquad (m_{H_2}^2 - (m_{H_1} -m_{h_0})^2)]^{1/2}
\end{eqnarray}
where   the step function is defined as $\theta(x) = 1 $  only for $x>0$  otherwise being zero.

\subsection{Relic density  analysis}
\begin{figure}[tb]
\begin{center}
\includegraphics[height=5.5 cm,  width=7.cm]{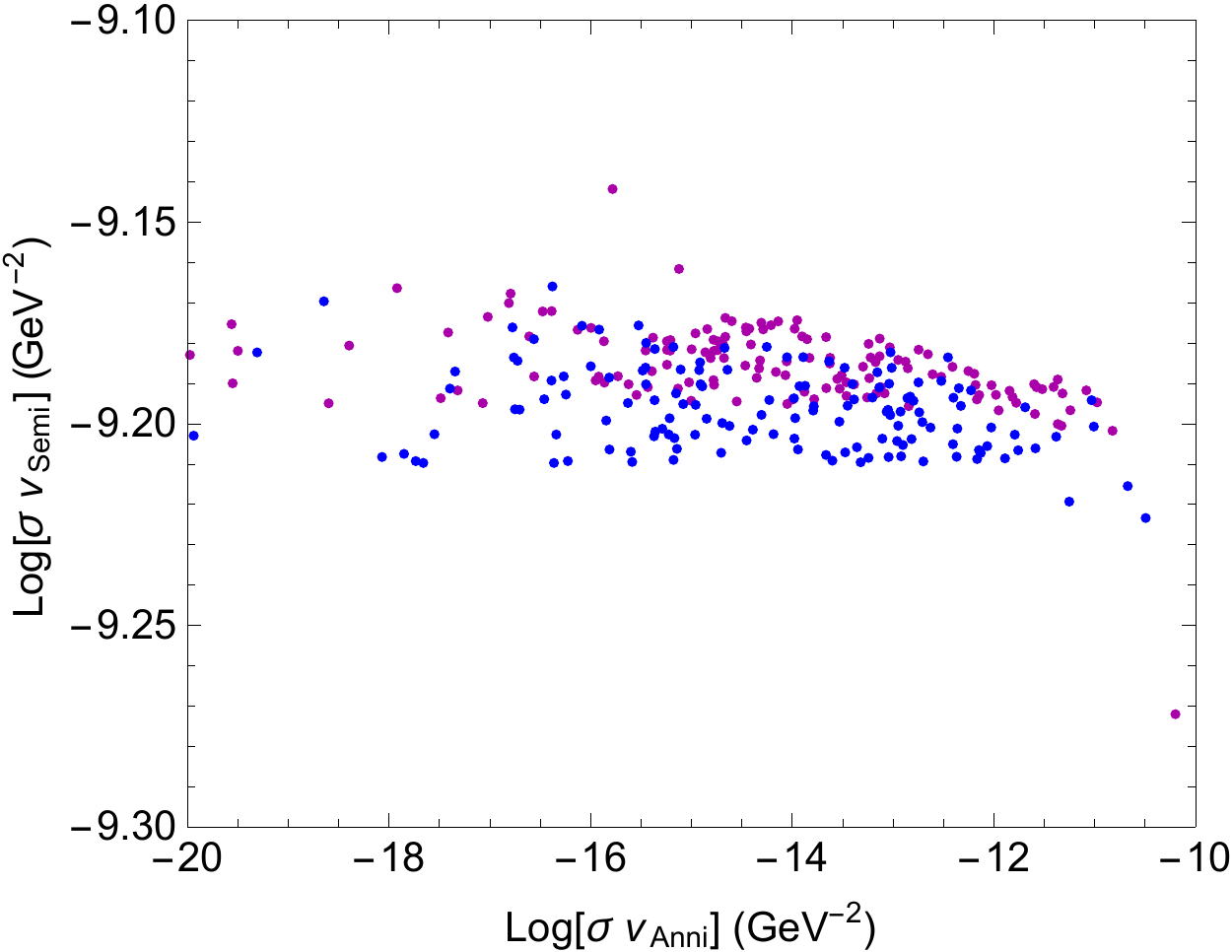} \qquad
\includegraphics[height=5.5 cm,  width=7.cm]{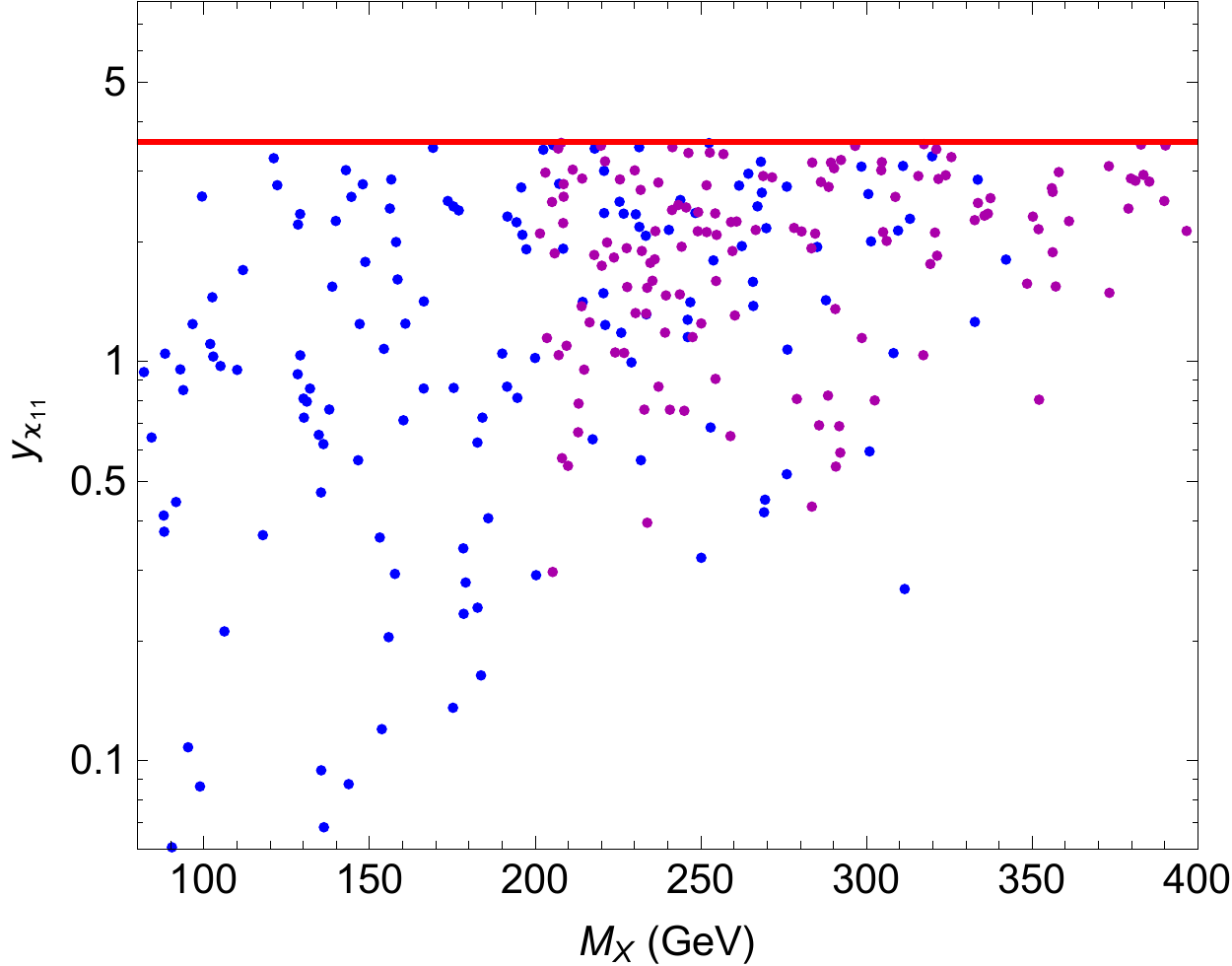}
\caption{The left plot shows the thermal average  $\langle \sigma v_{\rm Anni} \rangle $ for annihilation  versus the thermal average   $\langle \sigma v_{\rm Semi} \rangle$ for semi-annihilation   at the freeze out  temperature; The right plot illustrates the allowed region  in the plane of  $(M_X, y_{\chi_{11}} )$  with the red line signalling the perturbation limit $y_{\chi_{11}} < \sqrt{4 \pi} $. The  blue points represent  the scenario of $m_{H_1}= 2 M_X$ (lighter resonance) and  the magenta points stand  for the  scenario of  $m_{H_2} = 2 M_X$ (heavier resonance).  All points satisfy the LFV  bounds, neutrino data and  Planck satellite  measurement $ 0.117< \Omega h^2 < 0.123$  at 3 $\sigma$ confidential level. }
\label{results}
\end{center}
\end{figure}
In this section, we will show the  numerical analysis to satisfy all the constraints   discussed in Section~\ref{sec:model}. We find out that after imposing the LFV  bounds and neutrino oscillation data,  one  DM-neutrino-scalar coupling $y_{\eta_{i 1}}$ populates in the range of $(10^{-3}, 1.0)$,  so that the  annihilation process  in this model can not account for a correct relic density.  However  a large enhancement for $\langle\sigma v \rangle$ could be achieved if  the semi-annihilation  proceeds through a $S$-channel  and  in the vicinity of one narrow-width resonance. Since  Eq.~(\ref{eq:M3}) indicates  two  resonances of complex scalars are deconstructive,   one condition  $ 100 <(m_{H_2} - m_{H_1}) < 150$ GeV is imposed  in the analysis, with the upper limit from  the $3 \sigma$ EWPT fit at $\cos \alpha = \frac{1}{\sqrt{2}}$. Thus  for a given DM mass, only one resonance can effectively be on-shell. On the other hand,  we will require $m_{H'_1} \simeq m_{H'_2}$, i.e. quasi-degenerate,  in order to satisfy the neutrino oscillation data. This condition  can be easily fulfilled if we set the mixing term $ \lambda'_0 H^\dag\eta' s^{\prime*}\varphi^* $ to be tiny.  In order to simplify the analysis, we adopt several assumptions  as below:
\begin{align}
&  m_{\eta^\pm}=m_{H_2},  \quad   y'_{\eta'}=y_{\eta'},\quad y'_{s'}=y_{s'} \nn \\
&   s_\alpha =s_{\alpha'}= c_\alpha = c_{\alpha'}=\frac{1}{\sqrt2},
\end{align}
We set $m_{\eta^\pm}=m_{H_2}$  which is consistent with the  EWPT bound as shown in  Figure~\ref{fig:ST} and  $\ y'_{\eta'},\ y'_{s'} $ are taken to be diagonal matrices.  Under these assumptions, a numerical scan is conducted for the parameter space  by imposing the relevant neutrino and LFV bounds and  limiting the relic density to be $ 0.117< \Omega h^2 < 0.123$.  We explore the two on-shell scenarios in two  overlapping DM mass regions  with  $m_{H_1} = 2 M_X$ for $ 80 < M_X < 350$ GeV   and   $m_{H_2} = 2 M_X$ for  $200 <M_X < 400$ GeV.  Furthermore, in order to work well under the Breit-Wigner narrow width  prescription, we remove the points with $\rm{max}(\frac{\Gamma_{H_{1}}}{m_{H_1}}, \frac{\Gamma_{H_{2}}}{m_{H_2}})> 0.2$.  For the latter case of $m_{H_2}= 2 M_X$, we will  impose a smaller splitting  $(m_{H_2} -m_{H_1}) \simeq (115, 125)$ GeV, thus $m_{H_1} \gg m_X$. This condition will  ensure $\Gamma_{H_1} \ll m_{H_1}$ and avoid co-annihilation from scalars.  From the left plot in Fig:~\ref{results} we can see that  the observed relic density dominantly comes from the semi-annihilation.  At the time of freeze out $x_f \approx 21.0 $ (calculated by Eq.(\ref{eq:xf})), the thermal average of cross section is within the range of $5.98\times10^{-10}~{\rm GeV^{-2}}\lesssim\langle\sigma v_{Semi}\rangle\lesssim 8.83 \times 10^{-10}~{\rm GeV^{-2}}$, where the larger value  normally corresponds to a larger DM mass. In the right  plot we  show the allowed region in the $(M_X, y_{\chi_{11}})$ plan with other parameters randomly scanned. The plot demonstrates  that  a small DM mass $M_X < 200$ GeV is more sensitive to  the lighter $H_1 + i A_1$ resonance  and permits a DM  Yukawa coupling $ y_{\chi_{11}} \gtrsim 0.1$. However for  $M_X > 200$ GeV,  our fitting analysis indicates  a larger  DM  coupling $y_{\chi_{11}}  \gtrsim 0.5 $, which is close to the perturbative limit $\sqrt{4 \pi}$ regardless of  the lighter or heavier resonance  scenario.
\begin{figure}[tb]
\begin{center}
\includegraphics[height=5.05 cm,  width= 7.0 cm]{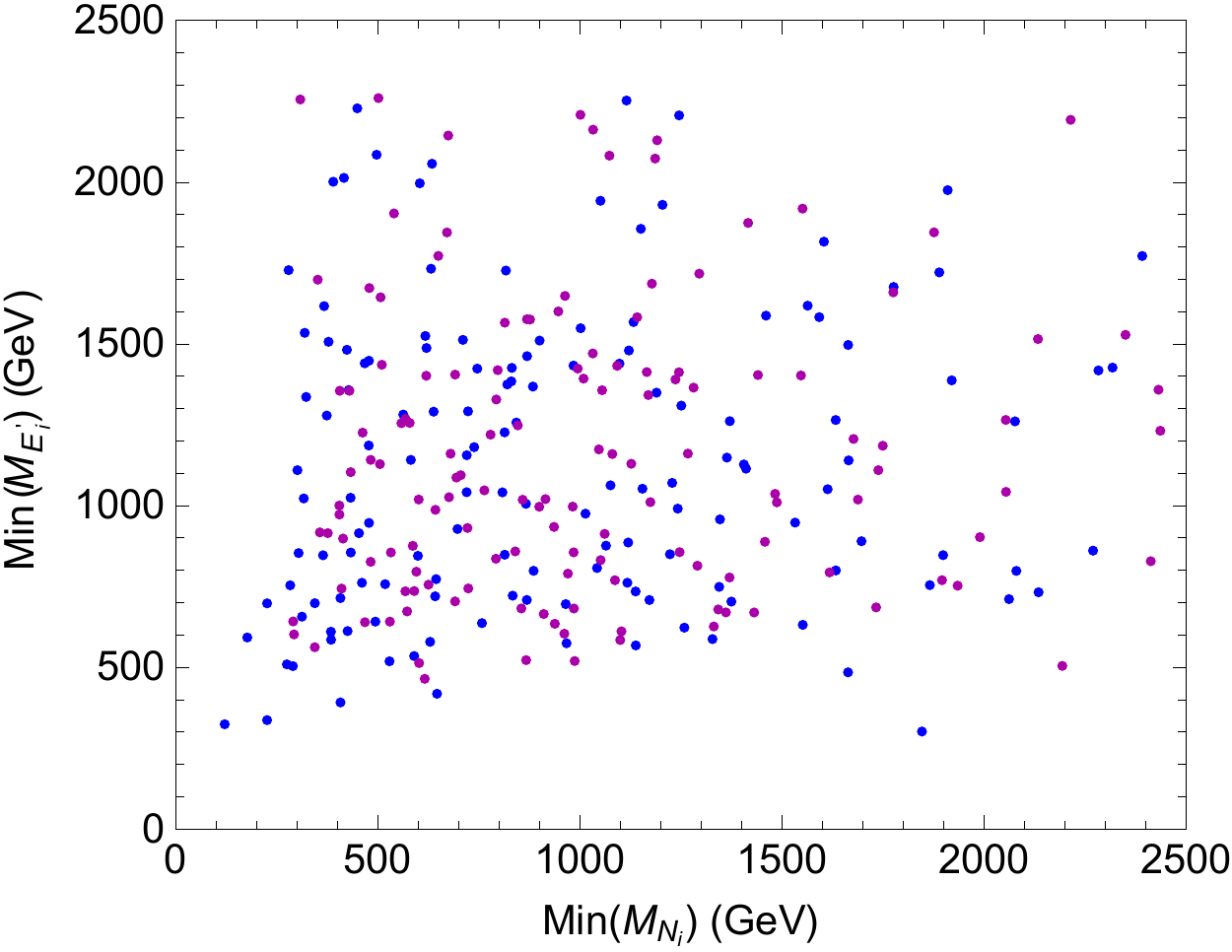} \qquad
\includegraphics[height=5.05 cm,  width= 7.0 cm]{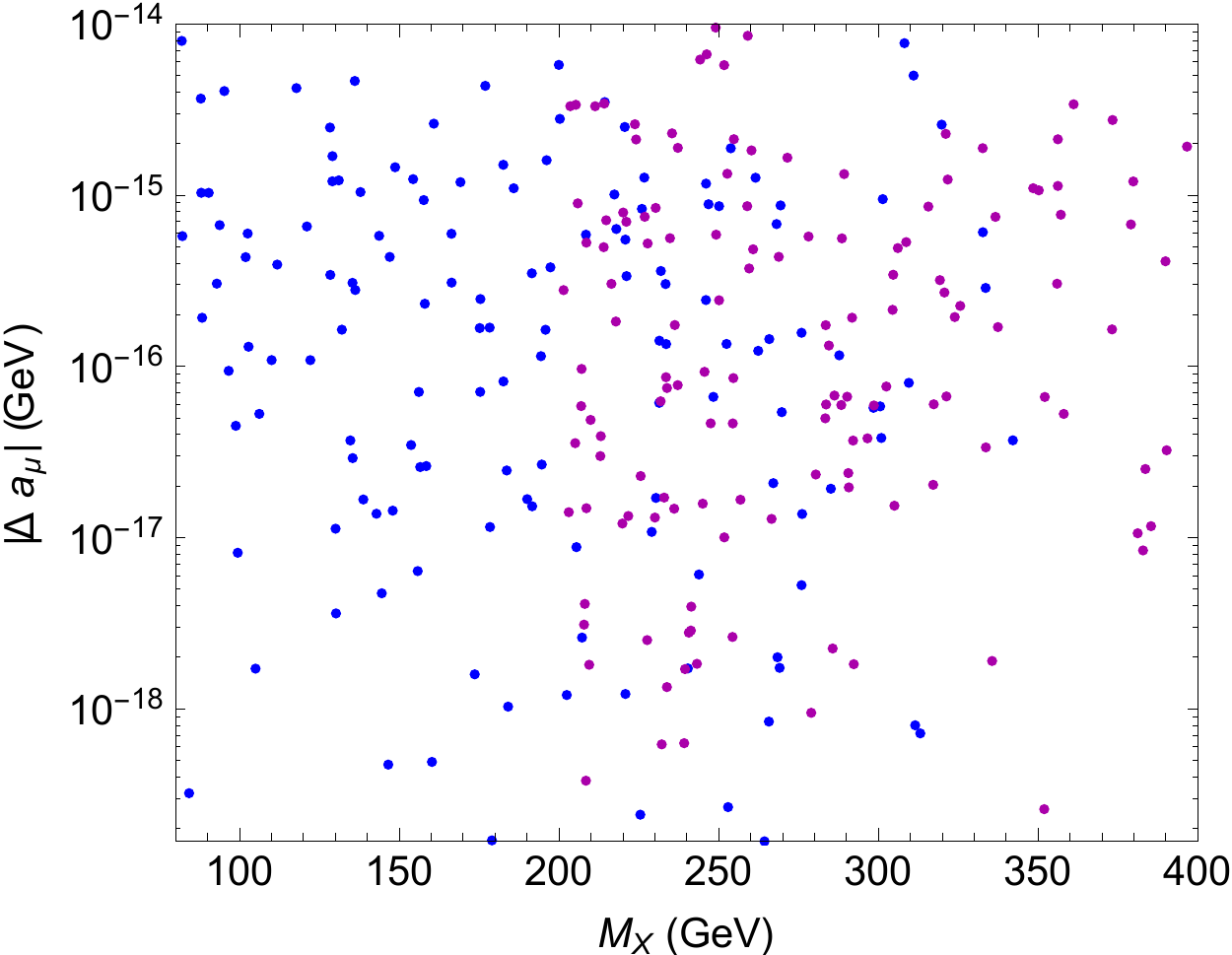}
\caption{The left plot shows the  lightest mass in  $M_{E'_i}, i =1,2,3$  versus the lightest mass in   $M_{N_i}, i= 1,2,3$; The right plot illustrates the correlation of $|\Delta a_\mu|$ to the DM mass $M_X$.  The  blue points represent  the scenario of  $m_{H_1}= 2 M_X$ and  the magenta points stand  for the scenario of  $m_{H_2} = 2 M_X$.  All points satisfy the LFV  bounds, neutrino data and  Planck satellite  measurement $ 0.117< \Omega h^2 < 0.123$  at 3 $\sigma$ confidential level. }
\label{sample}
\end{center}
\end{figure}

Fig.~\ref{sample} presents the mass ranges for  $M_{E'}(=M_{N'})$  and  $M_N$  which enter into the numerator of  neutrino mass form factors  as well as  values of $|\Delta a_\mu|$ versus $M_X$.  The typical value for the lightest vector-like fermions $L'_i$ lies in  $0.5 - 2.5$ TeV, but the degeneracy results in no effect on EWPT. Also this mass range of $M_{E'}$ is  not sensitive to the LHC bound for charged lepton pairs plus missing transverse energy~\cite{Cai:2018upp}.  While  after enforcing all the bounds, the maximum value for $|\Delta a_\mu|$  is of order $\lesssim 10^{-14}$,  even lower  for  most benchmark points, is negligible compared with the $3.6~\sigma$ deviation  of  order $10^{-9}$ as measured by the experiment.  Thus this model can not simultaneously account for the large discrepancy  in muon $g-2$.

{\it Direct detection}: 
In our case, there are no direction interactions between $H_{1,2}/A_{1,2}$  and  quarks at the tree level,
therefore  the constraints of direct detection searches should  be satisfied  without difficulty.

\section{ Conclusions and discussions}
We have constructed a neutrino mass model based on hidden local $U(1)_H$ symmetry which  gives rise to a  Dirac fermion type of Dark matter.  The neutrino masses are generated at the two-loop level due to  the symmetry and   particle content.  Furthermore because the form factor of the neutrino mass is proportional to the mass squared differences of  inert scalars, we  require one set of inert scalars to be quasi-degenerate so that  a sub-eV scale neutrino mass can be achieved without large fine-tuning for the Yukawa couplings. As  variation to this model, we illustrate that the heavy $Z'$ associated with the  $U(1)$  will  not impact  the DM annihilation because its mixing with SM $Z$ boson is induced by  a  complex triplet field $\Delta$,  whose VEV  is severely constrained by  $\rho$-parameter. Particularly,  the presence of  inert scalars $(\eta, s)$ gives rise to notable $S$ and $T$ deviations. Note that the impact of  singlet $s$ on oblique parameters is via the mixing with  doublet $\eta$. The $3 \sigma$ EWPT fit  prefers the mass splitting of  $|m_{H_2} -m_{H_1}| \lesssim 150$ GeV   provided  $\cos \alpha = \frac{1}{\sqrt{2}}$ and  $m_\eta^+ = m_{H_2}$.

Our DM is is the lightest neutral particle stabilised by  a discrete   $\mathbb{Z}_3$ parity which  is a residual  symmetry of $U(1)_H$ after spontaneous symmetry breaking.  Therefore, in addition to the standard DM annihilation process,  DM semi-annihilation is   induced in this model. After imposing the  LFV  bounds and neutrino oscillation data  and assuming no specific flavour structure in Yukawa couplings,  we find out that the $S-$channel semi-annihilation plays an important role to determine the observed relic density with a DM mass of $\mathcal{O}(100)$ GeV.  Our analysis demonstrates  that  the  lighter and heavier resonances  can  contribute significantly  when either  one  is actually put  on-shell  and  the  allowed  DM-scalar Yukawa coupling is in the range of   ($0.1$-$\sqrt{4 \pi}$) depending on the DM mass region.

\section*{Acknowledgments}

The research of H.C. is supported by the Ministry of Science, ICT and Future Planning of Korea, the Pohang City Government, and the Gyeongsangbuk-do Provincial Government.



\begin{appendix}
\section{Loop functions for neutrino mass} \label{app:neutrino}

The neutrino mass in this radiative seesaw model is generated by the two-loop Feynman diagrams in Figure~\ref{fig:neutrino}. It is convenient to  decompose the mass matrix  as $(m_{\nu})_{ij}
=m_{\nu_{ij}}^{(I)} + m_{\nu_{ij}}^{(II)} +[m_{\nu_{ij}}^{(I)}]^T + [m_{\nu_{ij}}^{(II)} ]^T$, with $m_{\nu_{ij}}^{(I)}$ and $m_{\nu_{ij}}^{(II)}$ calculated to be:
\begin{align}
m_\nu^{(I)} &= y_{\eta_{ia}} y_{s'_{a \rho }}^T y_{\eta'_{\rho b}} y_{S_{b j }}^T  s_{\alpha} c_{\alpha} s'_{\alpha} c'_{\alpha} \int\frac{d^4 k_1}{(2 \pi)^4} \int\frac{d^4 k_2}{(2 \pi)^4} \frac{- M_{N_\rho} k_2^2 }{(k_1^2 -M_{N_\rho}^2)(k_2^2 -M_{X_a}^2)(k_2^2 -M_{N'_b}^2)} \nonumber \\
& \left(\frac{1}{k_2^2 -m_{H_1}^2 } - \frac{1}{k_2^2 -m_{H_2}^2} \right) \left(\frac{1}{(k_1-k_2)^2 -m_{H'_1}^2 } - \frac{1}{k_2^2 -m_{H'_2}^2} \right) 
 \end{align}
 
 \begin{align}
m_\nu^{(II)} &= y_{\eta_{ia}} y_{s'_{a \rho}}^{\prime T} y'_{\eta'_{\rho b}} y_{S_{b j}}^T  s_{\alpha} c_{\alpha} s'_{\alpha} c'_{\alpha} \int\frac{d^4 k_1}{(2 \pi)^4} \int\frac{d^4 k_2}{(2 \pi)^4} \frac{M_{\chi_a} M_{N_\rho} M_{N'_{b}}}{(k_1^2 -M_{N_\rho}^2)(k_2^2 -M_{X_a}^2)(k_2^2 -M_{N'_b}^2)} \nonumber \\
& \left(\frac{1}{k_2^2 -m_{H_1}^2 } - \frac{1}{k_2^2 -m_{H_2}^2} \right) \left(\frac{1}{(k_1-k_2)^2 -m_{H'_1}^2 } - \frac{1}{k_2^2 -m_{H'_2}^2} \right) 
 \end{align}
For clarity, we can redefine $m_{\nu_{ij}}^{(I/II)} = \frac{1}{(4 \pi)^4} ~y_{\eta_{i a}} F_{I/II}(H_{1,2}, H'_{1,2})_{ab} ~ y_{b j}^T$ by extracting out a loop factor and Yukawa couplings in the outer loop of Feynman diagrams. After imposing  the Feynman parametrisation, the two loop functions $F_{I/II} (H_{1,2}, H'_{1,2})$   are given by:
\begin{align}
& F_I (H_{1,2}, H'_{1,2})_{a b}=  2 ~ y^T_{s'_{a\rho}} M_{N_\rho} y_{\eta'_{\rho b}}
(m^2_{H_1}-m^2_{H_2})(m'^2_{H_1}-m'^2_{H_2})s_{\alpha} c_{\alpha} s_{\alpha'} c_{\alpha'}\times\nn\\
&\int\frac{[da]_3 [d\alpha]_5 ~ a ( b+c) }
{[\alpha(a ~M^2_{N_\rho} + b~ m^2_{H'_1}+c~ m^2_{H'_2}) + a(b+c)(\beta ~M^2_{\chi_a} + \gamma ~M^2_{N'_b} +\rho ~ m^2_{H_1}+\sigma ~m_{H_2}^2)]^2},  \\  \nn \\
& F_{II}(H_{1,2}, H'_{1,2})_{a b} =  2 ~ M_{\chi_a}y'^T_{s'_{a\rho}} M_{N_\rho} y'_{\eta'_{\rho b}} M_{N'_b}
(m^2_{H_1} - m^2_{H_2}) (m'^2_{H_1} - m'^2_{H_2}) s_{\alpha} c_{\alpha} s_{\alpha'} c_{\alpha'}\times\nn\\
&\int\frac{[da]_3 [d\alpha]_5 ~ a^2 ( b+c)^2}
{[\alpha(a ~M^2_{N_\rho} + b~ m^2_{H'_1}+c~ m^2_{H'_2}) + a(b+c)(\beta ~M^2_{\chi_a} + \gamma ~M^2_{N'_b} +\rho ~ m^2_{H_1}+\sigma ~m_{H_2}^2)]^3} ,
\end{align}
where we use the definitions: $[da]_3 \equiv \int_0^1db\int_0^{1-b}dc$ with $a=1-b-c$, and $[d\alpha]_5 \equiv \int_0^1d\alpha\int_0^{1-\alpha}d\beta\int_0^{1-\alpha-\beta}d\gamma \int_0^{1-\alpha-\beta -\gamma} d\rho$ with $\sigma=1-\alpha-\beta-\gamma-\rho$. Note that these form factors are finite and  will be numerically evaluated.

\section{T parameter from mixing inert scalars} \label{app:Tpara}

\begin{figure}[tb]
\begin{center}
\includegraphics[scale=0.8]{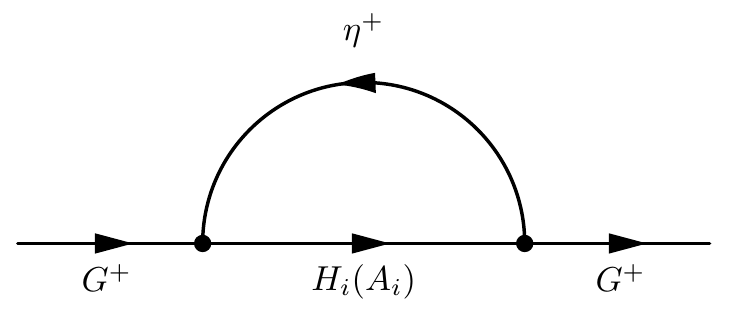} \quad \quad
\includegraphics[scale=0.8]{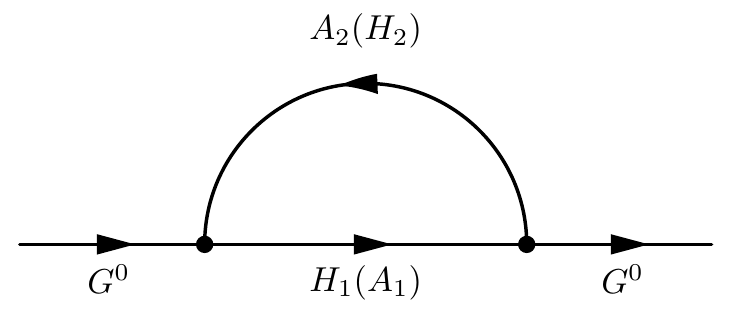}
\caption{ Self-energy diagrams for wave-function renormalisation.} 
\label{fig:rho}
\end{center}
\end{figure}

Since the longitude modes of  $W, Z$ gauge bosons are  $\partial_{\mu} G^{\pm, 0}$, the $T$ parameter is easily calculated from the wave-function renormalisation of  Goldstone bosons. We are going to  show that two approaches are matching with each other.  The relevant terms from the scalar potential  are:

\begin{align}
\mathcal{V} &\supset -\mu_H^2 H^\dagger H + \lambda_H ( H^\dagger H)^2 +\lambda_{H\eta} (H^\dag H) (\eta^\dag \eta ) + \lambda'_{H\eta} (H^\dag \eta) (\eta^\dag H ) +\lambda_{\varphi\eta} (\varphi^\dag H) (\eta^\dag \eta ) \nn \\
& + \lambda_{Hs} (H^\dag H) (s^* s ) +   \lambda_{\varphi s} (\varphi^\dag \varphi) (s^* s ) + (\lambda_0 H^\dag \eta s^* \varphi + h.c.) + \mu_\eta^2 \eta^\dag \eta + \mu_s^2 s^* s \label{eq:pot}
\end{align} 
Due to  the $\mathbb{Z}^3$ parity, there is no mass splitting among the imaginary and real parts of inert neutral scalars. The masses can be read off from Eq.(\ref{eq:pot}):
\begin{eqnarray}
& &m_{\eta^+}^2 = \mu_{\eta}^2 + \frac{1}{2} \left( \lambda_{H\eta} v_H^2 + \lambda_{\varphi \eta} v_{\varphi}^2 \right) \label{eq:m+}\\
& &  \frac{1}{ 2 }\begin{pmatrix}  s_{R/I} \\ \eta_{R/I} \end{pmatrix}^T  
\left[\begin{array}{cc}
m_{s_R}^2 &  \frac{1}{2} \lambda_0 v_H v_\varphi  \\ 
\frac{1}{2} \lambda_0 v_H v_\varphi & m_{\eta_{R}}^2 \\ 
\end{array}\right] \begin{pmatrix}  s_{R/I} \\ \eta_{R/I} \end{pmatrix} 
\end{eqnarray}
with the diagonal parts to be
\begin{eqnarray}
&&  m_{s_R}^2 = \mu_s^2 + \frac{1}{2} \left( \lambda_{H s} v_H^2 + \lambda_{\varphi s} v_{\varphi}^2 \right)  \nn \\
 && m_{\eta_R}^2 = \mu_{\eta}^2 + \frac{1}{2} \left( \lambda_{H\eta} v_H^2 + \lambda_{\varphi \eta} v_{\varphi}^2  + \lambda_{H \eta}^\prime v_H^2 \right)  \label{eq:m0}
\end{eqnarray}
The following identities will hold for the mass eigenstates and rotating angle:
\begin{eqnarray}
m_{s_R}^2 &=& m_{H_1}^2 \cos^2 \alpha + m_{H_2}^2 \sin^2 \alpha \nn \\
m_{\eta_R}^2 &=& m_{H_1}^2 \sin^2 \alpha + m_{H_2}^2 \cos^2 \alpha \nn \\
\sin 2 \alpha &=& \frac{\lambda_0 v_H v_\varphi}{m_{H_1}^2- m_{H_2}^2} \label{eq:alpha}
\end{eqnarray}
since  $ \delta \rho = \delta Z_{G^+} -  \delta Z_{G^0}$,  the two-point self-energy diagrams in Fig.~\ref{fig:rho} give us: 
\begin{eqnarray}
\hat{\alpha} \Delta T &=&  2 (\lambda_{H \eta}^\prime \frac{v_H}{2} \sin \alpha + \lambda_0 \frac{v_\varphi}{2} \cos \alpha)^2 f(m_{H_1}, m_{\eta^+}) \nn \\ &+& 2  (\lambda_{H \eta}^\prime \frac{v_H}{2} \cos \alpha - \lambda_0 \frac{v_\varphi}{2} \sin \alpha)^2 f(m_{H_2}, m_{\eta^+}) \nn \\
& - & \frac{1}{2} \lambda_0^2 v_\varphi^2 f(m_{H_1}, m_{H_2})  \label{eq:rho}
\end{eqnarray}
with the function $f(m_1, m_2) = - i \frac{d \Pi (p^2)}{d p^2}|_ {p^2 =0}$, and $\hat{\alpha} = \frac{e^2}{4 \pi}$. The  $\Pi (p^2)$  is defined as:
\begin{eqnarray}
\Pi (p^2) &=& \int  \frac{d^4 k}{ (2 \pi)^4}  \frac{1}{(k^2 - m_1^2)((k+p)^2 -m_2^2)} \nn \\
&=&   \int  \frac{d^4 k}{ (2 \pi)^4} \int_0^1 dx \frac{1}{(k^2 -\Delta)^2}
\end{eqnarray}
with $\Delta = - p^2 (1-x) x + x m_1^2 + (1-x) m _2^2 $. Thus we can obtain the analytic formula:
\begin{eqnarray}
f(m_1, m_2 ) &=&  \frac{1}{16 \pi^2} \frac{m_1^4 - m_2^4 +2 m_1^2 m_2^2 \log
   \left(\frac{m_2^2}{m_1^2}\right)}{2 (m_1^2- m_2^2)^3} \label{eq:fm}
\end{eqnarray}
Using Eqs.(\ref{eq:m+}), (\ref{eq:m0}), (\ref{eq:alpha}),  the coefficients in Eq.(\ref{eq:rho}) are related to be:
\begin{eqnarray}
&& (\lambda_{H \eta}^\prime \frac{v_H}{2} \sin \alpha + \lambda_0 \frac{v_\varphi}{2} \cos \alpha)^2  = \frac{(m_{H_1}^2 - m_{\eta^+}^2)^2}{v_H^2} \sin^2 \alpha  \nn \\
&& (\lambda_{H \eta}^\prime \frac{v_H}{2} \cos \alpha - \lambda_0 \frac{v_\varphi}{2} \sin \alpha )^2 = \frac{(m_{H_2}^2 - m_{\eta^+}^2)^2}{v_H^2} \cos^2 \alpha  \nn \\
&&  \lambda_0^2 v_\varphi^2 = \frac{4 (m_{H_1}^2 - m_{H_2}^2)^2}{v_H^2} \sin \alpha^2 \cos^2 \alpha  \label{eq:cc}
\end{eqnarray}
Then after substituting those identities back to Eq.(\ref{eq:rho}),  we obtain  the  $\Delta T$  expression in Eq.(\ref{eq:ST}).

\section{Inner products for the amplitudes} \label{app:formula}

\vspace{-0.5 cm}

\begin{eqnarray}
&& pk=\sqrt{((s-m_1^2-m_2^2)^2-4m_1^2m_2^2)((s-n_1^2-n_2^2)^2-4n_1^2n_2^2)}, \nn \\
&& p_1\cdot k_1 = \frac{1}{4s}\left(|(s+m_1^2-m_2^2)(s+n_1^2-n_2^2)|-pk \cos\theta \right), \nn \\
&& p_1\cdot k_2 = \frac{1}{4s}\left(|(s+m_1^2-m_2^2)(s+n_2^2-n_1^2)|+pk \cos\theta \right), \nn \\
&& p_2\cdot k_1 =\frac{1}{4s}\left(|(s+m_2^2-m_1^2)(s+n_1^2-n_2^2)|+pk \cos\theta \right), \nn \\
&& p_2\cdot k_2 =\frac{1}{4s}\left(|(s+m_2^2-m_1^2)(s+n_2^2-n_1^2)|-pk \cos\theta \right).
\end{eqnarray}
where $s\equiv (p_1+p_2)^2$ is a Mandelstam valuable, $m_{1,2}(p_{1,2})$ are initial state masses(momenta), while $n_{1,2}(k_{1,2})$ are final state masses(momenta).

\end{appendix}

\end{document}